\documentclass[twocolumn,epjc3]{svjour3}
\usepackage{graphicx}
\usepackage{amsfonts}
\usepackage{amsmath}
\usepackage{amssymb}
\usepackage[numbers,sort&compress]{natbib}
\usepackage[colorlinks,linkcolor=red,anchorcolor=blue,citecolor=green]{hyperref}
\usepackage{subcaption}
\usepackage{float}
\usepackage{enumerate}
\usepackage{multirow}
\newcommand\diff{\mathrm{d}}

\newcommand\e{\mathrm{e}}

\journalname{Eur. Phys. J. C}
\begin{document}
\title{
    Realistic neutron star models in $f(T)$ gravity
}
\author{
    Rui-Hui Lin\thanksref{e1,addr1}
    \and
    Xiao-Ning Chen\thanksref{addr1}
    \and
    Xiang-Hua Zhai\thanksref{e2,addr1}
}

\thankstext{e1}{e-mail: linrh@shnu.edu.cn}
\thankstext{e2}{e-mail: zhaixh@shnu.edu.cn}
\institute{Division of Mathematics and Theoretical Physics, Shanghai Normal University,\\ 100 Guilin Road, Shanghai 200234, China \label{addr1}}
\maketitle
\begin{abstract}
    We investigate the nonrotating neutron stars in $f(T)$ gravity with $f(T)=T+\alpha{T}^2$,
    where $T$ is the torsion scalar in the teleparallel formalism of gravity.
    In particular, we utilize the SLy and BSk family of equations of state for perfect fluid to describe the neutron stellar matter
    and search for the effects of the $f(T)$ modification on the models of neutron stars.
    For positive $\alpha$, the modification results in a smaller stellar mass in comparison to general relativity,
    while the neutron stars will contain larger amount of matter for negative $\alpha$.
    Moreover, there seems to be an upper limit for the central density of the neutron stars with $\alpha>0$,
    beyond which the effective $f(T)$ fluid would have a steplike phase transition in density and pressure profiles,
    collapsing the numerical system.
    We obtain the mass-radius relations of the realistic models of neutron stars and subject them to the joint constraints
    from the observed massive pulsars PSR J0030+0451, PSR J0740+6620, and PSR J2215+5135,
    and gravitational wave events GW170817 and GW190814.
    For the neutron star model in $f(T)$ gravity to be able to accommodate all the mentioned data,
    the model parameter $\alpha$ needs to be smaller than $-4.295$, $-6.476$, $-4.4$, and $-2.12$
    (in the unit of ${G}^2M_\odot^2/c^4$) for SLy, BSk19, BSk20, and BSk21 equations of state, respectively.
    If one considers the unknown compact object in the event GW190814 not to be a neutron star
    and hence excludes this dataset,
    the constraints can be loosened to
    $\alpha<-0.594$, $-3.5$, $0.4$ and $1.9$ (in the unit of ${G}^2M_\odot^2/c^4$), respectively.

\end{abstract}

\section{Introduction}
\label{intro}
General relativity (GR) seems to work perfectly well
against the local weak field tests of gravity.
However, the long known challenge of it when applied to the entire Universe,
including the dark contents of the Universe
and the singularities of the spacetime,
still lacks a consensus solution.
It is believed that a quantum theory of gravitation may possibly help understand or resolve these problems.
Nonetheless, since a generally accepted theory of quantum gravity is still missing,
modifications of GR which may hint the quantum corrections are widely considered.
In this sense, the Einstein-Hilbert action of GR may be seen as a classical approximation at low energy scale.
In the Riemannian formulation of GR, gravitation manifests itself in the curvature of the spacetime manifold.
One would expect that higher order curvature term(s) may become relevant as the energy goes higher.
Therefore, alternative theories of gravity involving nonlinear terms of curvature appear to be of particular interests,
among which $f(R)$ model with an arbitrary function $f$ of the Ricci scalar $R$ is one of the most renowned schemes
(see, e.g., Refs. \cite{DeFelice:2010aj,Sotiriou:2008rp,Capozziello:2011et,Nojiri:2010wj} for extensive reviews).

On the other hand, GR can also be formulated in various forms by choosing different affine connections,
which may constitute different but equivalent descriptions of gravity
\cite{BeltranJimenez:2019tjy,PhysRevD.101.024053} and provide different aspects of insight.
As one of the variants of GR that can be dated back to the time of Einstein,
the teleparallel equivalent of GR (TEGR) can be formulated
where the tetrad field is used as the dynamical variable instead of the metric
and the torsion scalar $T$ is constructed to be the underlying Lagrangian while the curvature $R$ vanishes
\cite{Aldrovandi:2013wha,Maluf2013}.
Following this line, the correction corresponding to higher energy scale may appear as terms of higher orders of torsion.
Similar to $f(R)$ gravity, this may be achieved by $f(T)$ gravity with a nonlinear function of $T$
(see, e.g., Refs.\cite{PhysRevD.79.124019,PhysRevD.81.127301,Cai:2015emx,Nojiri:2017ncd}).
Although TEGR is equivalent to GR,
the extended theories, $f(R)$ and $f(T)$, are generally different in that
the difference between their Lagrangians
is no longer a total derivative term when the function $f$ is nonlinear,
and hence cannot be discarded via integration by parts.
It follows that $f(T)$ gravity may possess new features and provide new angle to investigate the geometry of the spacetime.
One particular advantage of $f(T)$ gravity is that
the field equations under this framework are of second order of derivatives as in GR
instead of fourth order ones in $f(R)$ gravity.
This makes $f(T)$ gravity somewhat a simpler and more natural way to modify (TE)GR.

However, in some early formulations of $f(T)$ gravity,
it was found that the Lorentz invariance is not respected by the theory
in the sense that the torsion tensor, and hence the field equations,
are in general not invariant with respect to the tetrad \cite{Sotiriou2011,Li2011,Ferraro2015}.
The frame-dependent nature of this formalism complicates the studies of various subjects in $f(T)$ gravity
and hinders the development of the theory,
which has then led to efforts in finding the \textit{good} tetrad in certain situations
\cite{FERRARO2011,Tamanini2012}.
But the head-on solution to this problem should be restoring the spin connection
and constructing the covariant formulation of teleparallel gravities
\cite{Obukhov2006,Krssak:2015oua,Golovnev2017,Lin:2019tos}.
This covariant formulation then allows one to pursue the relatively more difficult problems
beyond Minkowski spacetime or spatial flat cosmology in $f(T)$ gravity,
which, for example, include the spherically symmetric configurations \cite{PhysRevD.94.124025,Golovnev:2021htv,Pfeifer:2021njm,DeBenedictis:2022sja}.
In particular, by considering the regularity of the matter,
the relativistic stars in $f(T)$ gravity are proved existing for \textit{a priori} assumed metrics\cite{Boehmer:2011gw}.
After that, various sorts of star solutions in $f(T)$ gravity have been found and studied in this approach
\cite{Zubair:2015cpa,NewtonSingh:2019bbm,Saha:2019msh,Nashed:2020kjh}.

Another route to study relativistic stars is to start from the physical properties of the matter that may form the stars.
Following this scheme, stars consisting of Yang-Mills field \cite{DeBenedictis:2018wkp} and boson field \cite{Ilijic:2020vzu} are investigated.
For stellar matter that can be approximated by perfect fluid,
if energy is conserved, a correlation between the spacetime metric and the stellar structure can be found.
In GR, this is the well-known Tolman-Oppenheimer-Volkoff (TOV) equation.
Moreover, the equation of state (EOS) of matter is also essential to the interior structure of the star.
Mathematically speaking, this relation between matter density and pressure is required to close the differential equation system.
Physically, EOS represents the model one employs to describe the stellar matter.
For a compact star like a neutron star (NS) in which the matter is so dense that a fully relativistic treatment is needed,
the interior structure is mostly supported by the degeneracy pressure of nuclear particles.
If one assumes a neutron stellar matter that involves pure nucleon-nucleon interaction,
a polytropic approximation of the EOS,
\begin{equation}
    \label{polytrope}
    p=\kappa \rho^\gamma,
\end{equation}
can be utilized,
where $p$ and $\rho$ are the pressure and density of the matter, respectively,
and $\kappa$ and $\gamma$ are parameters related to the matter model.
The compact stars in $f(T)$ gravity utilizing the polytropic EOS has been studied in Ref. \cite{PhysRevD.98.064047}.
In reality, however, NSs
may consist of several layers that have different physical properties.
A unified EOS that is valid in all of these crusts and core segments is then needed to construct realistic models of stars.
Various unified EOSs have been employed to study the stellar structures in $f(R)$ gravity
\cite{Arapoglu:2010rz,Astashenok:2013vza,Capozziello:2015yza,Astashenok:2017dpo}.
A comprehensive review on the relativistic stars in modified gravities can be found in Ref. \cite{Olmo:2019flu}.

In addition to the theoretical development of NS models,
recent observations suggest that massive NSs with the mass close or beyond the prediction of GR may exist\cite{Rhoades:1974fn}.
In particular, from the astrophysical data released by various projects including Neutron Star Interior Composition Explorer (NICER),
North American Nanohertz Observatory for Gravitational Waves (NANOGrav) and Green Bank Telescope (GBT),
it can be derived that there exist massive pulsars with masses in the range $1.44\sim 2.27M_\odot$ \cite{Linares:2018ppq,Miller:2019cac,NANOGrav:2019jur}.
Moreover, the gravitational wave (GW) events of binary mergers
observed by LIGO/Virgo collaboration also provide independent measurement
of the properties of the event participants,
among which the GW signal GW190814 indicates that a compact object, possibly an NS, with a mass around $2.6M_\odot$ may exist \cite{Abbott:2020khf}.
The existence of these super-massive NSs may be explained by the modification of gravitational theories such as $f(T)$ gravity studied in this work.
And hence, it is necessary to compare the theoretical predictions with the observational data and put constraints to the theory
so that it can accommodate the observed NSs.

In this paper, we intend to study the realistic models of NSs in $f(T)$ gravity.
We aim at obtaining the stellar structure numerically and searching for possible signatures of the modification of $f(T)$.
We use the simple but realistic model $f(T)=T+\alpha T^2$ that can be seen as an analog to the Starobinsky model in curvature-based gravity\cite{Starobinsky:1980te}.
The quadratic term of torsion scalar, like its curvature counterpart, can describe the early time cosmic inflation \cite{Nashed:2014vsa,Rezazadeh:2015dza}.
It is then reasonable to consider NSs in such a model in that the extreme gravitational environment within an NS may be similar to the condition of the early universe
and hence be an excellent natural laboratory to study the gravity in such a regime.
Analogic consideration has been given to the study of NSs in realistic $f(R)$ models that also describe inflation \cite{Astashenok:2013vza,Capozziello:2015yza,Astashenok:2017dpo}.
Moreover, this quadratic model of $f(T)$ gravity has also been reconstructed to avoid finite-time future singularities \cite{Bamba:2012vg} and alleviate the Hubble tension problem \cite{Cai:2019bdh}.
Within this model, we employ several unified EOSs (SLy and BSk family) to describe the neutron stellar matter.
Then the NSs in $f(T)$ gravity are subjected to comparison with
the observations of massive pulsars and GW events.

The paper is organized as follows.
In Sec. \ref{tele}, we briefly review the basis of the covariant teleparallel gravities
and set the equations for spherically symmetric stellar structure.
Concrete models and the internal structures of NSs in $f(T)$ gravity are presented numerically in Sec. \ref{struc}.
Based on the numerical results,
we present the mass-radius relations of the NSs and subject them to the constraints of observations in Sec. \ref{mrs}.
Section \ref{conclusion} contains our concise summary and discussions.
Throughout the paper, we use the units with $c=8\pi G=1$.

\section{Equations for stelllar structure in covariant $f(T)$ gravity}
\label{tele}
\subsection{The covariant $f(T)$ gravity}
As the spacetime manifold $\mathcal M$ is assumed to be a parallelizable metric space,
one can generally find a trivialization $e_a=e_a^{\:\mu}\partial_\mu$ of the tangent bundle of the manifold.
The dual vector basis 1-form to $e_a$, i.e. the tetrad, is given by $h^a=h^a_{\:\mu}\diff x^\mu$,
so that $h^a(e_b)=\delta^a_b$.
The spacetime metric $g$ is related to the tangent space metric $\eta$ by
\begin{equation}
    \label{geta1}
    g=g_{\alpha\beta}\diff x^\alpha\otimes\diff x^\beta=\eta_{ab}h^a\otimes h^b,
\end{equation}
or, in terms of components,
\begin{equation}
    \label{geta2}
    g_{\alpha\beta}=\eta_{ab}h^a_{\:\alpha} h^b_{\:\beta},\quad \eta_{ab}=g_{\alpha\beta}e_a^{\:\alpha}e_b^{\:\beta}.
\end{equation}
The torsion 2-form is given by\cite{Aldrovandi:2013wha,Krssak:2015oua}
\begin{equation}
    T^a=\mathcal D h^a=\diff h^a+\omega^a_{\:b}\wedge h^b,
    \label{covariantT}
\end{equation}
where the covariant exterior derivative $\mathcal D$ and the spin connection $\omega^a_{\:b}$ are introduced
such that for any vector $V^a$ in the tangent space at a given point,
$\mathcal D_\mu V^a$ is covariant under Lorentz rotation.
Generally, one can always find a specific tetrad, called proper tetrad,
in which all components of the spin connection vanish.
Hence, in terms of components, the torsion tensor with proper tetrad is
\begin{equation}
    \label{torsionts}
    T^\alpha_{\:\beta\gamma}=e_a^{\:\alpha}\left( \partial_\beta h^a_{\:\gamma}-\partial_\gamma h^a_{\:\beta} \right).
\end{equation}
Then, the torsion scalar is given by
\begin{equation}
    T=T^a\wedge\star\left( T_a-h^a\wedge\left( e_b\cdot T^b \right)-\frac12 e^a\cdot\left( h^b\wedge T_b \right) \right)
    \label{exttorsion}
\end{equation}
where $\star$ denotes the Hodge dual and $\cdot$ indicates the interior product.
In terms of contraction of tensors, it is
\begin{equation}
    \label{torsionsc}
    T=T^\alpha_{\:\beta\gamma}S_\alpha^{\:\beta\gamma},
\end{equation}
with the super potential
\begin{equation}
    \label{sp}
    S_\alpha^{\:\beta\gamma}=\frac14 \left( T_\alpha^{\:\beta\gamma}+T^{\gamma\beta}_{\:\;\:\;\alpha}-T^{\beta\gamma}_{\:\;\:\;\alpha} \right)+\frac12 \left( \delta^\beta_\alpha T^{\lambda\gamma}_{\:\;\:\;\lambda}-\delta^\gamma_\alpha T^{\lambda\beta}_{\:\;\:\;\lambda} \right).
\end{equation}
TEGR takes $T$ as its Lagrangian,
and the $f(T)$ gravity considers an arbitrary function of $T$ instead, i.e.,
\begin{equation}
    \label{ftlag}
    \mathcal{S}=-\frac12\int|h|f(T)\diff^4 x+\int|h|\mathcal{L}_M\diff^4x,
\end{equation}
where $|h|=\det(h^a_{\alpha})=\sqrt{-g}$ is the determinant of the tetrad $h^a_{\alpha}$,
and $\mathcal{L}_M$ is the Lagrangian of matter.
Variation of Eq. \eqref{ftlag} with respect to the tetrad gives the field equations
\begin{equation}
    \label{fieldeq}
    \frac2{|h|}\partial_\beta \left( |h|S_\sigma^{\:\alpha\beta}e_a^{\:\sigma}f_T \right)+\frac f2 e_a^{\:\alpha}=\mathcal{T}_\beta^\alpha e_a^{\:\beta},
\end{equation}
where $f_T$ denotes $\diff f/\diff T$,
and the energy-momentum tensor $\mathcal{T}_\beta^\alpha$ of matter is given by
\begin{equation}
    \label{emts}
    \frac{\delta(|h|\mathcal{L}_M)}{\delta h^a_{\:\alpha}}=|h|\mathcal{T}_\beta^\alpha e_a^{\:\beta}.
\end{equation}

\subsection{Spherically symmetric stellar equations}
For spherically symmetric stars,
we consider a static metric
\begin{equation}
    \label{sphmetric}
    \diff s^2=\e^{A(r)}\diff t^2-\e^{B(r)}\diff r^2-r^2\diff\theta^2-r^2\sin^2\theta\diff\phi^2,
\end{equation}
where the unknown metric functions $A(r)$ and $B(r)$ depend only on the radial coordinate $r$.
The choice of the tetrad for the spherically symmetric configuration in teleparallel framework is an issue that has been paid much attention to.
In the early-time formulations of $f(T)$ gravity,
the spin connection $\omega^a_{\;b}$ is set to be zero as a priori regardless of the choice of tetrad,
which leads to a frame-dependent nature of the gravitation theory \cite{Sotiriou2011,Li2011,Ferraro2015}.
Some efforts have been devoted to find out the suitable tetrad for specific scenarios \cite{FERRARO2011,Tamanini2012}.
However, finding a \textit{good} tetrad for every symmetry does not remove the theory's dependence on the choice of reference.
In fact, the spin connection $\omega^a_{\;b}$ that relates to the reference transformation cannot be always zero.
Restoring the spin connection associated with each frame is the key to recover the covariance of the $f(T)$ gravity
\cite{Obukhov2006,Krssak:2015oua,Golovnev2017,Lin:2019tos}.
In this covariant framework, any tetrad satisfying Eq. \eqref{geta2} is eligible,
as long as the corresponding spin connections are restored.
Furthermore, since the spin connection relates to the frame transformation,
it is always possible to find a frame in which all components of $\omega^a_{\;b}$ vanish
and the torsion can be simply written in the form \eqref{torsionts}.
This tetrad is called the proper tetrad, which,
unlike the \textit{good} tetrads for specific symmetries, is not a preferred frame,
but rather a choice of convenience.
Therefore, the calculations and results based on the proper tetrad are physical
in that they respect the covariance of the theory and do not depend on the choice of reference.
In the following, we continue our discussions using the proper tetrad.
As discussed in Refs. \cite{Obukhov2006,Krssak:2015oua,Golovnev2017}, the proper tetrad suitable for the metric \eqref{sphmetric} can be chosen as
\begin{equation}
    \label{tetrad}
    h^a_{\:\alpha}=\left(
    \begin{array}{cccc}
            \e^{\frac A2} & 0                               & 0                               & 0                       \\
            0             & \e^{\frac B2}\sin\theta\cos\phi & \e^{\frac B2}\sin\theta\sin\phi & \e^{\frac B2}\cos\theta \\
            0             & -r\cos\theta\cos\phi            & -r\sin\theta\sin\phi            & r\sin\theta             \\
            0             & r\sin\theta\sin\phi             & -r\sin\theta\cos\phi            & 0
        \end{array}
    \right).
\end{equation}
Then, the torsion scalar is given by
\begin{equation}
    \label{torsionscsph}
    T(r)=\frac2{r^2}\e^{-B(r)}\left( \e^{\frac{B(r)}{2}}-1 \right)\left( \e^{\frac{B(r)}{2}}-1-rA'(r) \right).
\end{equation}
The stellar matter, if considered as a perfect fluid, can be described by
\begin{equation}
    \label{perfectfluid}
    \mathcal{T}_{\mu\nu}=(p+\rho)u_\mu u_\nu-pg_{\mu\nu}
\end{equation}
with the 4-velocity $u^\mu$.
The $tt$ and $rr$ components of Eq.\eqref{emts} read
\begin{equation}
    \label{eomsph}
    \begin{split}
        \rho=&-\frac{\e^{-B}}{2r^2}\left\{ 4rf_{TT}T'\left( \e^{\frac B2} -1 \right)\right.\\
        &\left.+2f_T \left[ rB'+\left( \e^\frac B2-1 \right)\left( 2+rA' \right) \right]+fr^2\e^B \right\},\\
        p=&\frac{f}2+\frac{2f_T\e^{-\frac B2}}{r^2}\left( 1-\e^{-\frac B2} \right)-\frac{f_T\e^{-\frac B2}}r \left( 2\e^{-\frac B2}-1 \right)A',
    \end{split}
\end{equation}
where the prime indicates a derivative with respect to $r$ and $f_{TT}=\diff^2f/\diff T^2$.
For a third equation, one can take either the angular component of Eq. \eqref{emts} or the conservation of matter $\nabla_\mu\mathcal{T}^{\mu\nu}=0$,
which gives the TOV equation
\begin{equation}
    \label{TOV}
    p'=-\frac12(\rho+p)A'.
\end{equation}

By extracting the Einstein tensor $G_{\mu\nu}$, Eq. \eqref{fieldeq} can be written as
\begin{equation}
    \label{efffieldeq}
    G_{\mu\nu}=\mathcal T_{\mu\nu}+\tilde{\mathcal{T}}_{\mu\nu},
\end{equation}
where the modification from the nonlinear term(s) of $f(T)$ is absorbed in the effective energy-momentum tensor $\tilde{\mathcal{T}}_{\mu\nu}$ of $f(T)$ \textit{fluid} given by
\begin{equation}
    \label{effemtensor}
    \tilde{\mathcal{T}}_{\mu\nu}=\left( f_T T-f \right)\frac12g_{\mu\nu}-2S_{\mu\nu}^{\:\;\:\;\sigma}\partial_\sigma f_T+\left( 1-f_T \right)G_{\mu\nu}.
\end{equation}
The components of Eq.\eqref{effemtensor} can be denoted as the effective density, radial and transverse pressures as follows
\begin{equation}
    \label{effrhop}
    \begin{split}
        \tilde{\rho}=&\frac f2+\frac{f_T}{r^2\e^B}\left[ \left( \e^{\frac B2}-1 \right)\left( 2+rA' \right)+rB' \right]\\
        &+\frac{2f_{TT}}{r\e^B}\left(\e^{\frac B2}-1\right)T'+\frac{1}{r^2\e^B}\left( \e^B+rB'-1 \right),\\
        \tilde p_r=&-\frac f2-\frac1{r^2\e^B}\left( 1-\e^B+rA' \right)\\
        &-\frac{f_T}{r^2\e^B}\left[ 2\left(\e^{\frac B2}-1\right)+\left( \e^{\frac B2}-2 \right)rA' \right],\\
        \tilde p_t=&-\frac f2+\frac{f_T}{4r^2\e^B}\left[ 4-8\e^{\frac B2}+4\e^B-2\left(2\e^{\frac B2}-3\right)rA'\right.\\
            &\left.+r^2A'^2-rB'\left(2+rA'\right)+2r^2A'' \right]\\
        &-\frac{f_{TT}}{2r\e^B}T'\left( 2\e^{\frac B2}-2-rA' \right)\\
        &-\frac1{4r\e^B}\left[ 2A'+rA'^2-B'\left(2+rA'\right)+2rA'' \right],
    \end{split}
\end{equation}
so that $\tilde{\mathcal{T}}_{\mu\nu}$ can be written effectively in the form of an anisotropic perfect fluid
\begin{equation}
    \label{aniso}
    \tilde{\mathcal{T}}_{\mu\nu}=(\tilde{\rho}+\tilde p_t)u_\mu u_\nu-\tilde p_t g_{\mu\nu}+(\tilde p_r-\tilde p_t)\chi_\mu\chi_\nu,
\end{equation}
where $\chi^\mu$ is the radial space-like unit vector.
We note here that the effective $f(T)$ \textit{fluid} is the extra part of the field equation when compared to the Einstein equation.
In the current work, we consider it merely the effect of the modifications from the $f(T)$ model
rather than an actual extra field.
In this sense, the difference between $\tilde p_t$ and $\tilde p_r$ only reflects the anisotropic nature of the field equation.

\section{Internal structure}
\label{struc}
In this section, we study numerically the internal structure of the NSs in $f(T)$ gravity.
As a simple model, we consider the concrete form of $f(T)=T+\alpha T^2$,
which can be viewed as an analog of the Starobinsky model in the curvature framework of gravity\cite{Starobinsky:1980te}
and may be a realistic model to describe inflationary universe \cite{Nashed:2014vsa,Rezazadeh:2015dza},
avoid future singularities \cite{Bamba:2012vg} and alleviate Hubble tension problem \cite{Cai:2019bdh}.
When $\alpha=0$, $f_T=1$ and $f_TT-f=0$,
one can easily check from Eq.\eqref{effemtensor} that $\tilde{\mathcal{T}}_{\mu\nu}=0$
and Eq.\eqref{efffieldeq} reduces to the Einstein equation in GR.
In this case, the NS model is identical to the GR model of NS.

\subsection{Boundary conditions and EOSs}
Since $A(r)$ does not appear explicitly in Eq. \eqref{eomsph} or \eqref{TOV},
we therefore solve the system for the functions $A'(r),B(r)$ and $\rho(r)$.
The boundary conditions appropriate for the system can be set at the center of the star.
The regularity condition then requires
\begin{equation}
    \label{bcs}
    A'(0)=0,\quad B(0)=0,\quad \rho(0)=\rho_c
\end{equation}
for some central density $\rho_c$ of the NS.
The surface of the star is generally defined at the radius where the radial pressure vanishes.
For self-bound model of stars, density does not always vanish here.

For the last piece to close the system given by Eqs. \eqref{eomsph} and \eqref{TOV},
one needs the EOS of the stellar matter, i.e., the algebra relation between $p$ and $\rho$.
Since the interaction of matter under the extreme environment within the NS is not yet very well understood,
various EOSs for the neutron stellar matter have been proposed to describe the different crust and core segments in a unifying way.
The main features of the representative EOSs that we use in the present paper are reported as follows:
\begin{itemize}
    \item From the Skyrme Lyon (SLy) effective nucleon-nucleon interaction,
          the SLy EOS for nonrotating neutron stellar matter is obtained by many-body methods \cite{Douchin:2001sv,Haensel:2004nu}.
          Nuclei in the neutron star crust are described by the compressible liquid drop model.
          The calculation is also continued to the characteristic densities of the expected liquid core of NSs.
          The analytic representation of SLy EOS can be written as
          \begin{equation}
              \label{sly}
              \begin{split}
                  \zeta=&\frac{a_1+a_2\xi+a_3\xi^3}{1+a_4\xi}k \left( a_5(\xi-a_6) \right)\\
                  &+\left( a_7+a_8\xi \right) k \left( a_9(a_{10}-\xi) \right)\\
                  &+\left( a_{11}+a_{12}\xi \right)k \left( a_{13}(a_{14}-\xi) \right)\\
                  &+\left( a_{15}+a_{16}\xi \right)k \left( a_{17}(a_{18}-\xi) \right),
              \end{split}
          \end{equation}
          where $\xi=\log_{10} \left( \rho/(\mathrm{g\:cm^{-3}}) \right)$ and $\zeta=\log_{10} \left( p/(\mathrm{dyn\:cm^{-2}}) \right)$,
          and the function $k(x)$ is defined as $k(x)=1/(\e^x+1)$.
    \item The BSk family\cite{Potekhin:2013qqa} of EOSs bases on the generalized Skyrme interaction, supplemented by several interaction corrections,
          which is developed for cold catalyzed nuclear matter.
          Each BSk EOS corresponds to a specific numerical fit.
          At the typical density of neutron matter,
          BSk19, BSk20, and BSk21 describe approximately the soft, moderate, and stiff matter EOS, respectively.
          The analytic representation of BSk EOSs can be written as
          \begin{equation}
              \label{bsk}
              \begin{split}
                  \zeta=&\frac{a_1+a_2\xi+a_3\xi^3}{1+a_4\xi}k \left( a_5(\xi-a_6) \right)\\
                  &+\left( a_7+a_8\xi \right) k \left( a_9(a_{6}-\xi) \right)\\
                  &+\left( a_{10}+a_{11}\xi \right)k \left( a_{12}(a_{13}-\xi) \right)\\
                  &+\left( a_{14}+a_{15}\xi \right)k \left( a_{16}(a_{17}-\xi) \right)\\
                  &+\frac{a_{18}}{1+\left( a_{19}(\xi-a_{20}) \right)^2}+\frac{a_{21}}{1+\left( a_{22}(\xi-a_{23}) \right)^2}.
              \end{split}
          \end{equation}

\end{itemize}
The parameters $a_i$ in Eqs. \eqref{sly} and \eqref{bsk} are listed in \ref{appeos}.
For these EOSs, vanishing points of pressure and density coincide.
Therefore, if the density profile decreases outwardly and reaches zero at some point,
the radius $\mathcal R$ of the star can be defined here, i.e.,
\begin{equation}
    \label{bc2}
    \rho(\mathcal R)=0.
\end{equation}

Note that for electromagnetic field and non-interacting matter,
the trace $\mathcal T$ of Eq. \eqref{perfectfluid} is positively determined, i.e., $\mathcal T=\rho-3p\ge0$.
This condition is then sometimes assumed to hold for any perfect fluid \cite{Capozziello:2015yza}.
If this is true, there will be an extra constraint on the density $\rho$ for the EOSs mentioned above.
However, it is pointed out that $\mathcal T$ may be negative for relativistic invariant, causal theories describing strong interacting systems \cite{Zeldovich:1961sbr}.
The correlation between the possibility of a negative $\mathcal T$ and the macroscopic properties of NSs is studied in Ref. \cite{Podkowka:2018gib}.
In the present work, we assume that $\mathcal T$ is allowed to be negative.

\begin{figure}[h!]
    \centering
    \includegraphics[width=.9\linewidth]{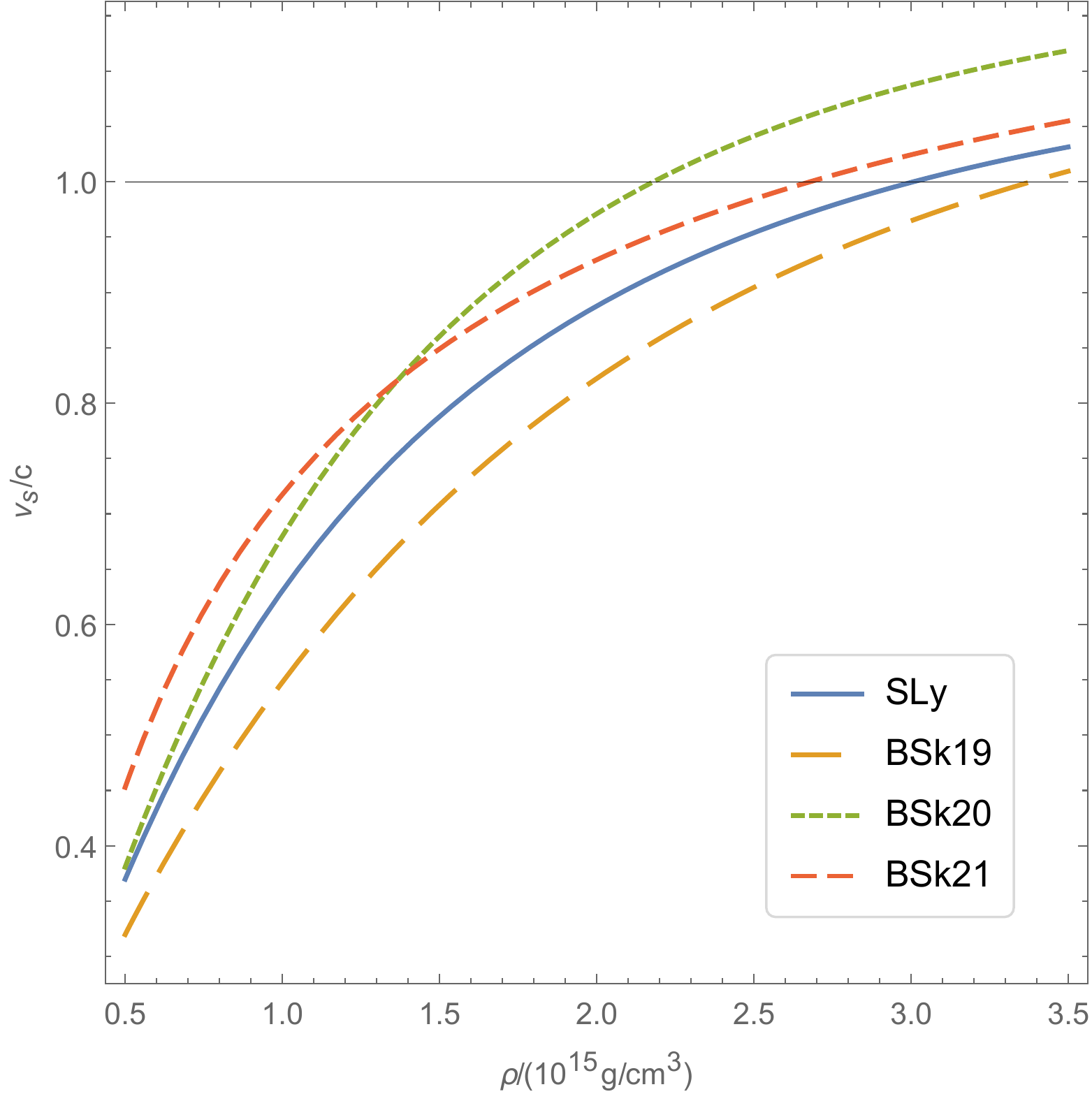}
    \caption{The speed of sound $v_s$ as a function of the stellar density $\rho$ for the EOSs SLy, BSk19, BSk20, and BSk21.
        The horizon line marks the causal limit $v_s=c$.}
    \label{cs}
\end{figure}

Furthermore, the speed of sound of the stellar matter
\begin{equation}
    \label{sos}
    v_s=\sqrt{\frac{\diff p}{\diff\rho}}
\end{equation}
should be less or equal to the speed of light
so that the causality will not be violated,
which may put a constraint on the density $\rho$.
For the EOSs mentioned above, $v_s$ can be expressed as a function of the density $\rho$, i.e.,
\begin{equation}
    \label{sos1}
    v_s(\rho)=\sqrt{\frac{p}{\rho}\frac{\partial \zeta}{\partial \xi}}=\sqrt{10^{\zeta-\xi}\frac{\partial \zeta}{\partial \xi}},
\end{equation}
for which the illustration is given in Fig. \ref{cs}.
One can see that the causal limit indeed puts constraints on $\rho$ for the EOSs considered.
And the maximum densities allowed are found to be $3.007,\:3.381,\:2.180,$ and $2.679$ (in the unit of $10^{15}\mathrm{g/cm^3}$)
for the EOSs SLy, BSk19, BSk20, and BSk21, respectively.
Beyond these limits, the speed of sound will be greater than the speed of light and the causality will be violated.

\begin{figure}[h!]
    \centering
    \includegraphics[width=.9\linewidth]{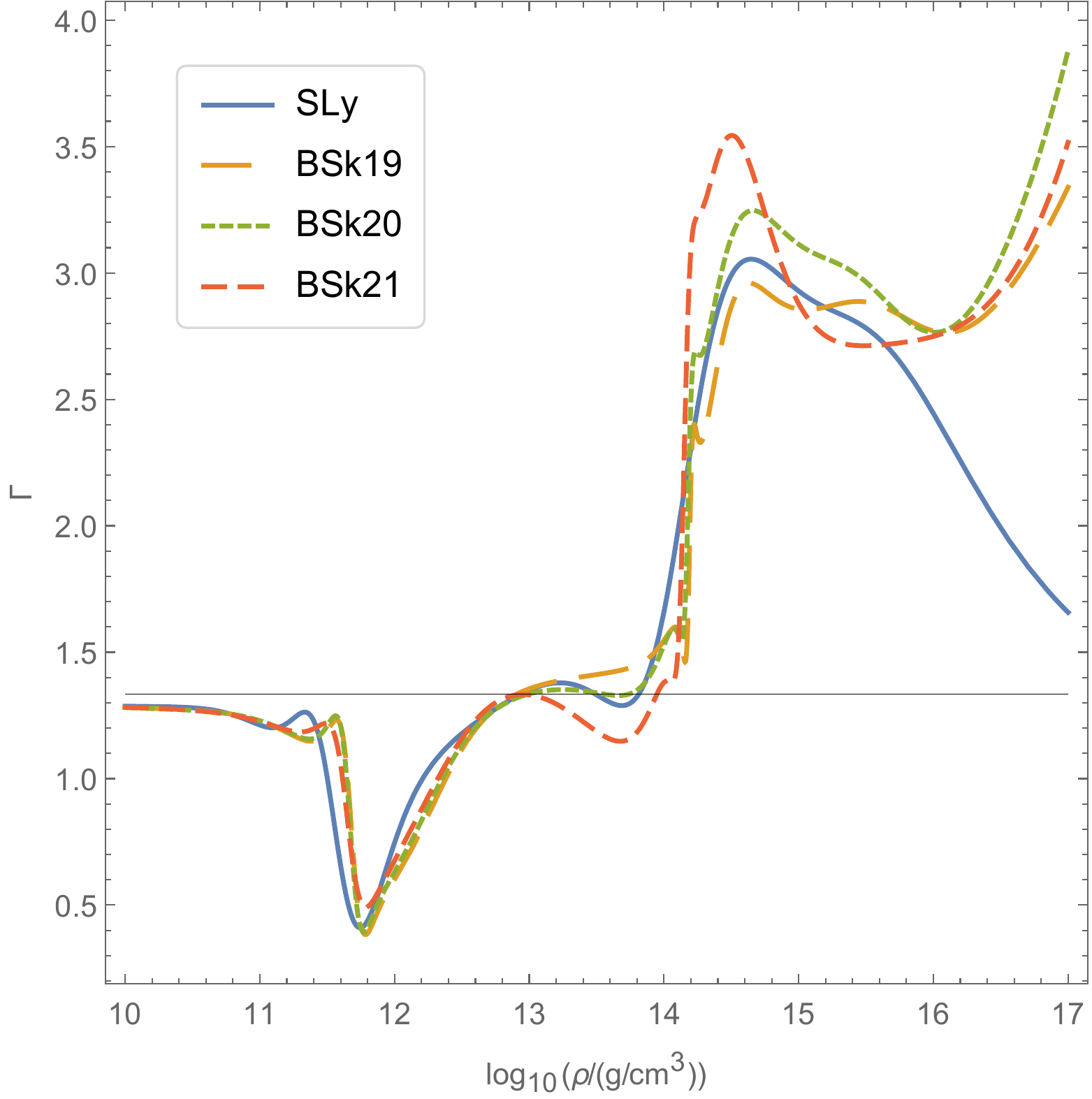}
    \caption{The adiabatic index $\Gamma$ for SLy and BSk EOSs.
        The horizon line marks $\Gamma=4/3$.}
    \label{gamma}
\end{figure}

Similarly, one can also express the adiabatic index as a function of the density $\rho$
\begin{equation}
    \label{adiabatic}
    \Gamma(\rho)=\frac{\rho+p}p \frac{\partial p}{\partial \rho}=\frac{v_s^2(\rho)}{c^2}+\frac{\partial \zeta}{\partial \xi},
\end{equation}
which, for the EOSs considered, are illustrated in Fig. \ref{gamma} (see also in Ref. \cite{Potekhin:2013qqa}).
One can see that for lower densities ($\leq 10^{13}\mathrm{g/cm^3}$),
the adiabatic index falls below the Chandrasekhar's condition $\Gamma\geq 4/3$.
This is because for these EOS models,
the pressure at this density is mainly determined by the pressure of ultrarelativistic electron gas.
Therefore, as long as the density profile decreases outwardly starting from some high enough value ($\gtrsim 10^{15}\mathrm{g/cm^3}$),
the interior of the neutron star can be divided into the stable, high density inner core
and the unstable outer crust.
The detail profiles of the stellar densities are discussed in the next subsection.

\subsection{Energy density and pressure profiles}
\begin{figure*}[h!]
    \centering
    \begin{subfigure}{0.49\linewidth}
        \includegraphics[width=\linewidth]{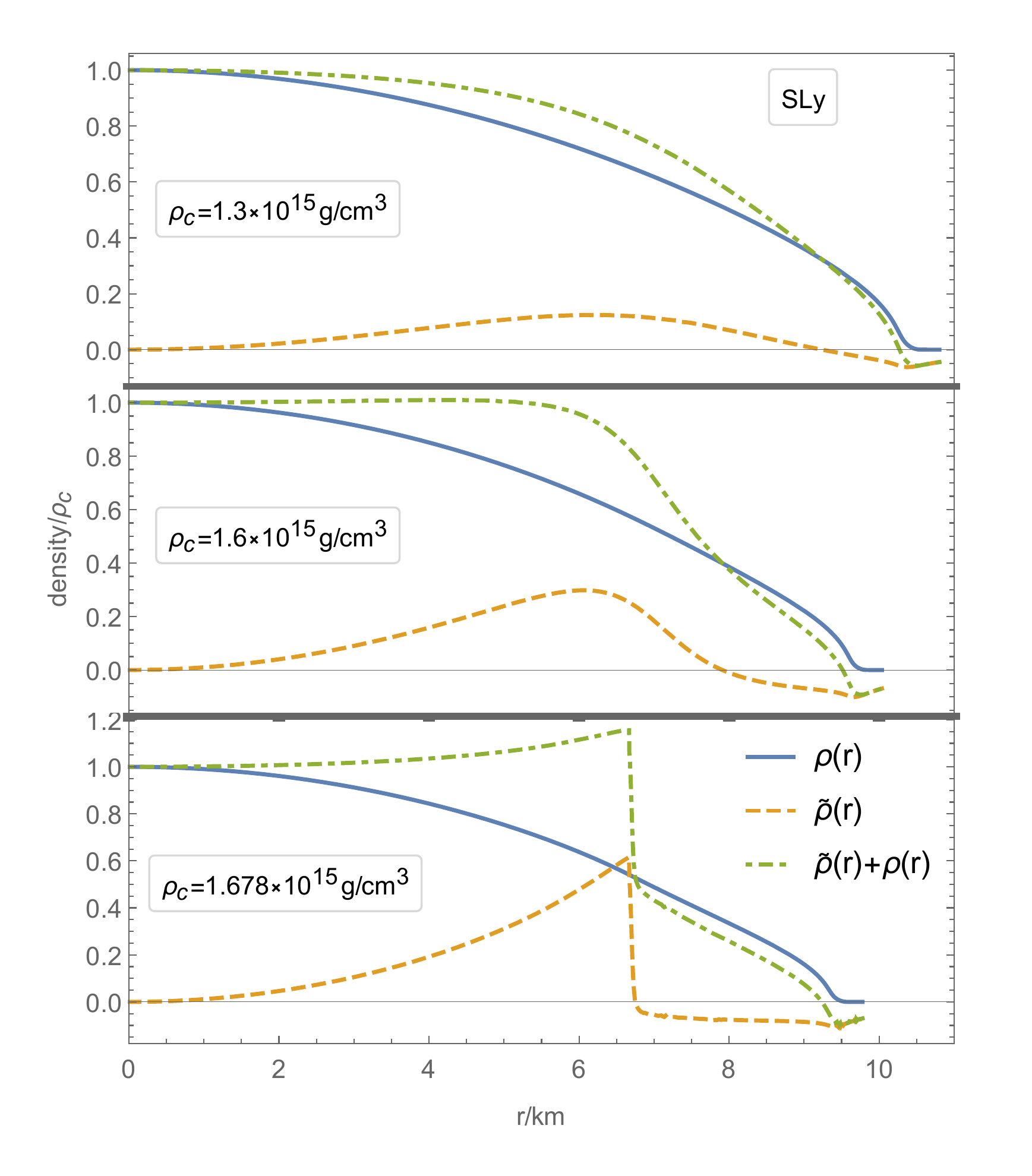}
    \end{subfigure}
    \begin{subfigure}{0.49\linewidth}
        \includegraphics[width=\linewidth]{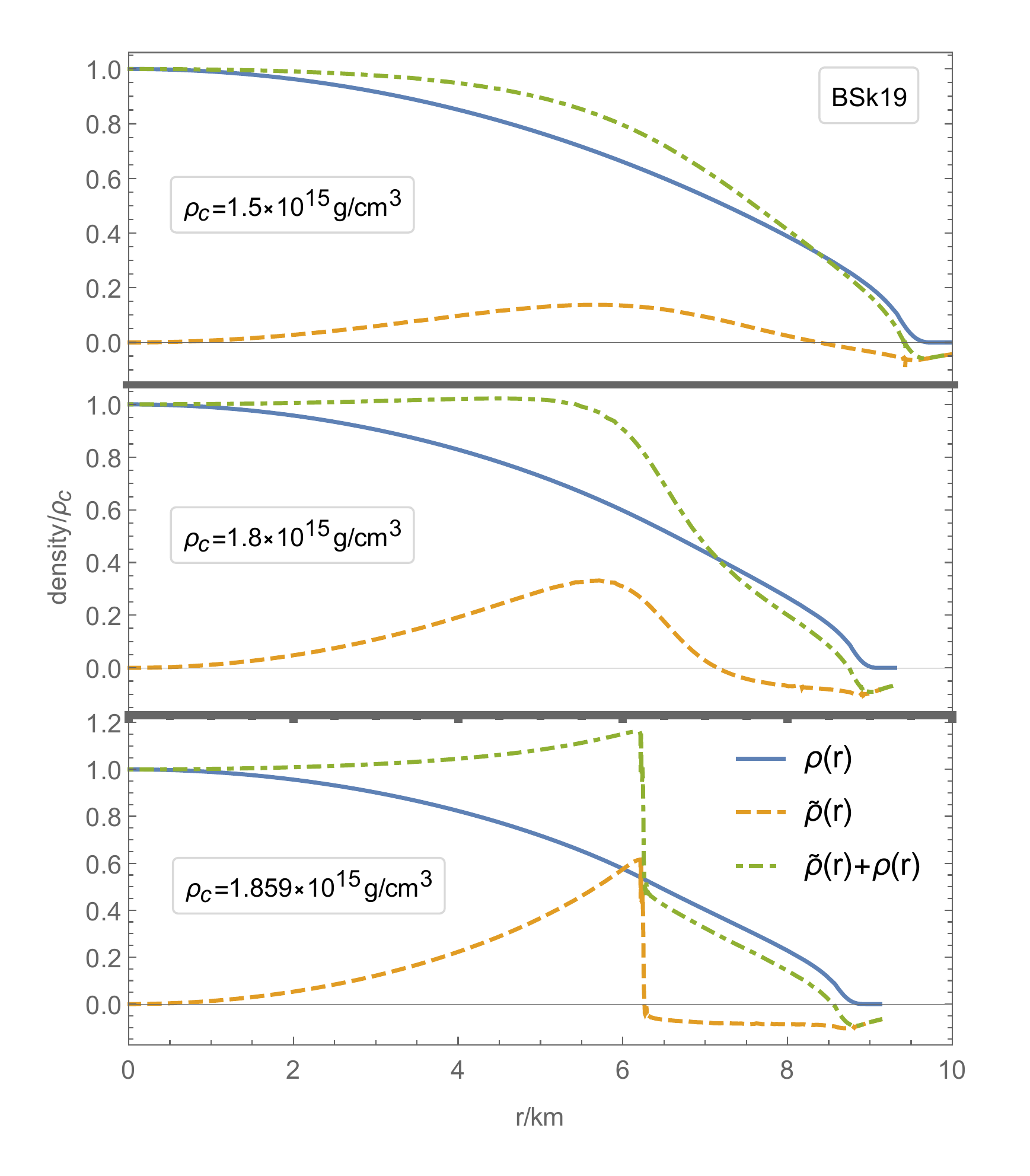}
    \end{subfigure}
    \begin{subfigure}{0.49\linewidth}
        \includegraphics[width=\linewidth]{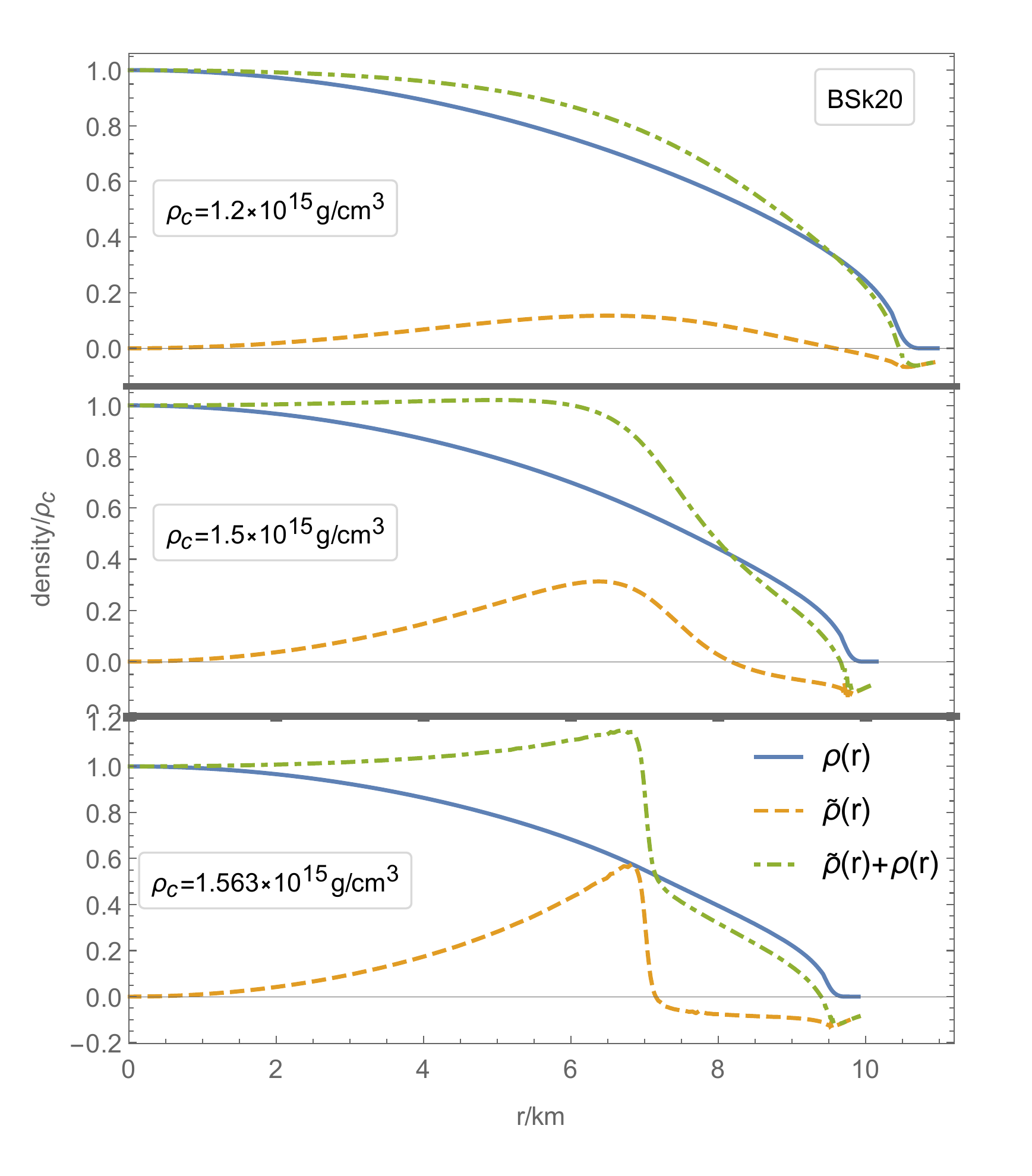}
    \end{subfigure}
    \begin{subfigure}{0.49\linewidth}
        \includegraphics[width=\linewidth]{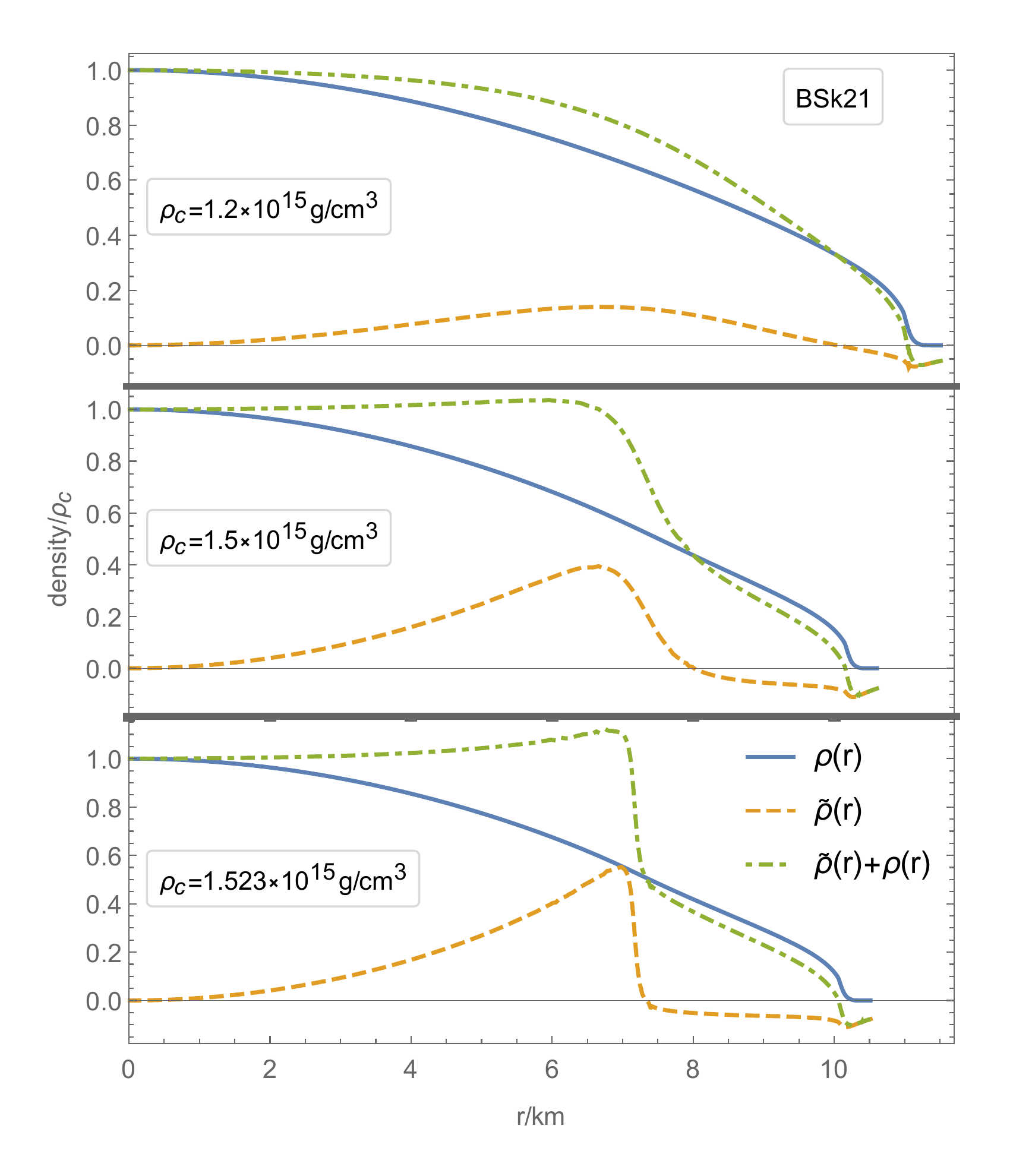}
    \end{subfigure}
    \caption{Density profiles, as well as the profiles of the effective density of the $f(T)$ \textit{fluid}, for SLy, BSk19, BSk20, and BSk21 EOSs
        with $\alpha=10r_g^2$ and several representative values of $\rho_c$.}
    \label{effrho_plus}
\end{figure*}

\begin{figure*}[h!]
    \centering
    \begin{subfigure}{0.49\linewidth}
        \includegraphics[width=\linewidth]{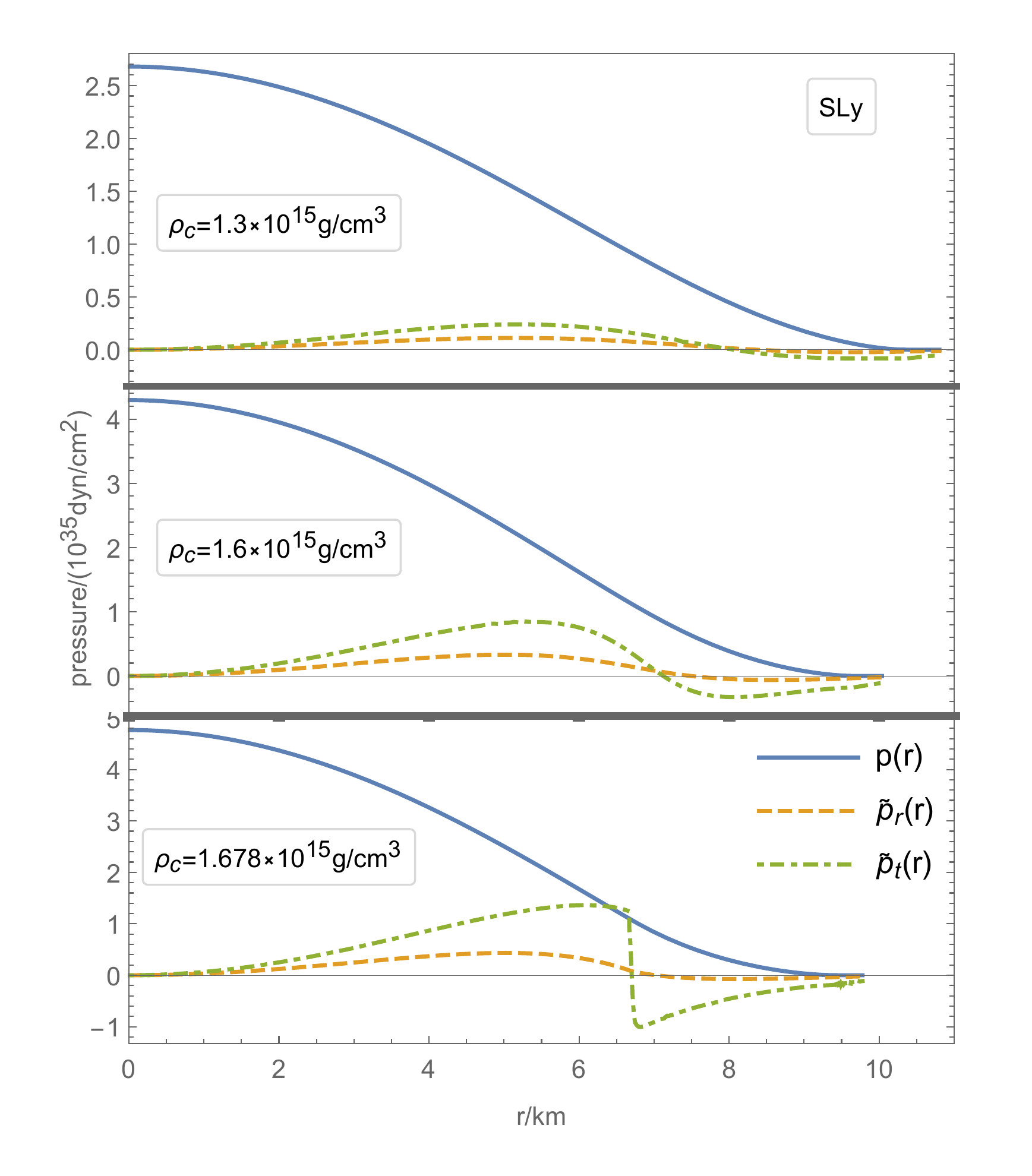}
    \end{subfigure}
    \begin{subfigure}{0.49\linewidth}
        \includegraphics[width=\linewidth]{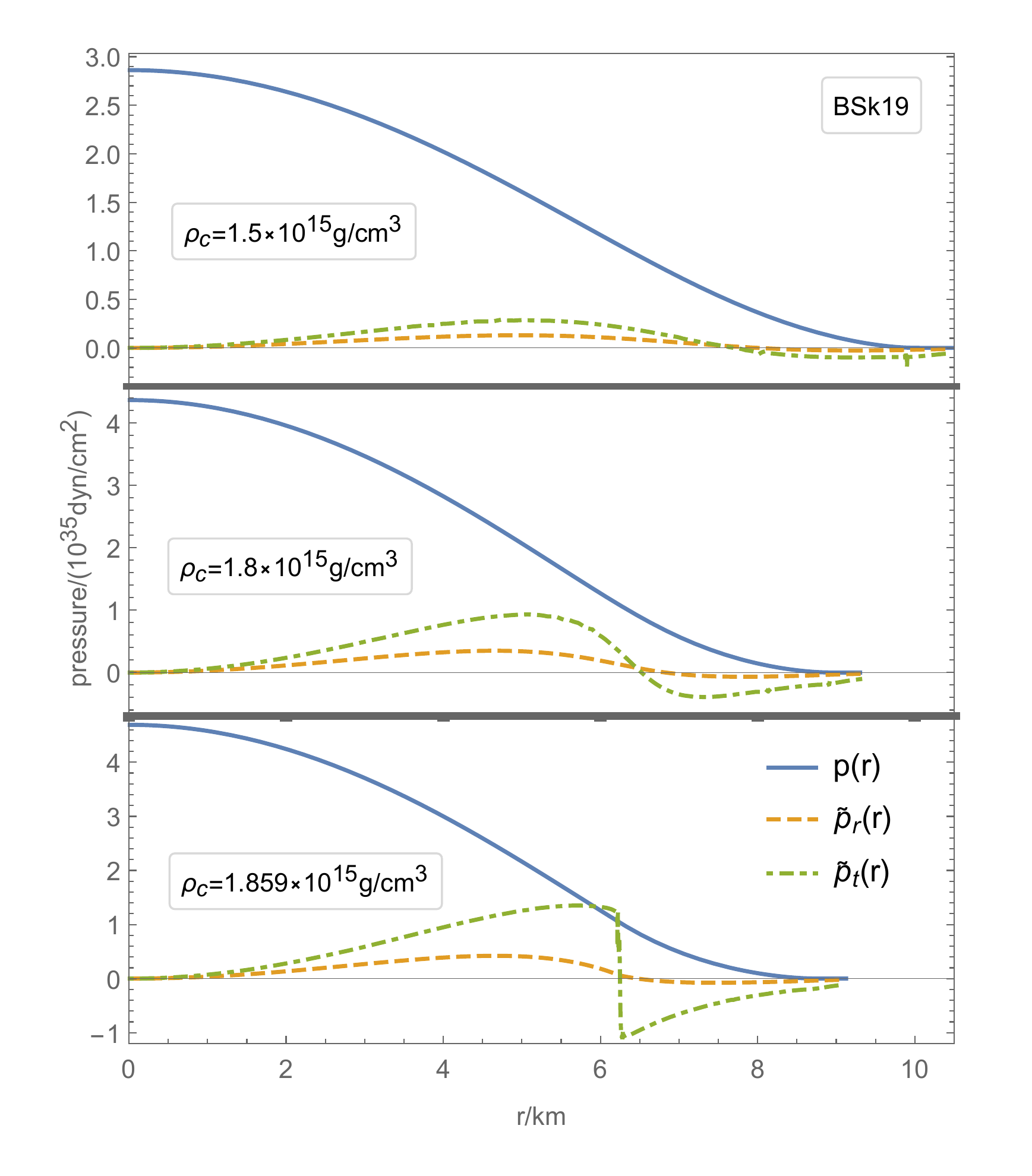}
    \end{subfigure}
    \begin{subfigure}{0.49\linewidth}
        \includegraphics[width=\linewidth]{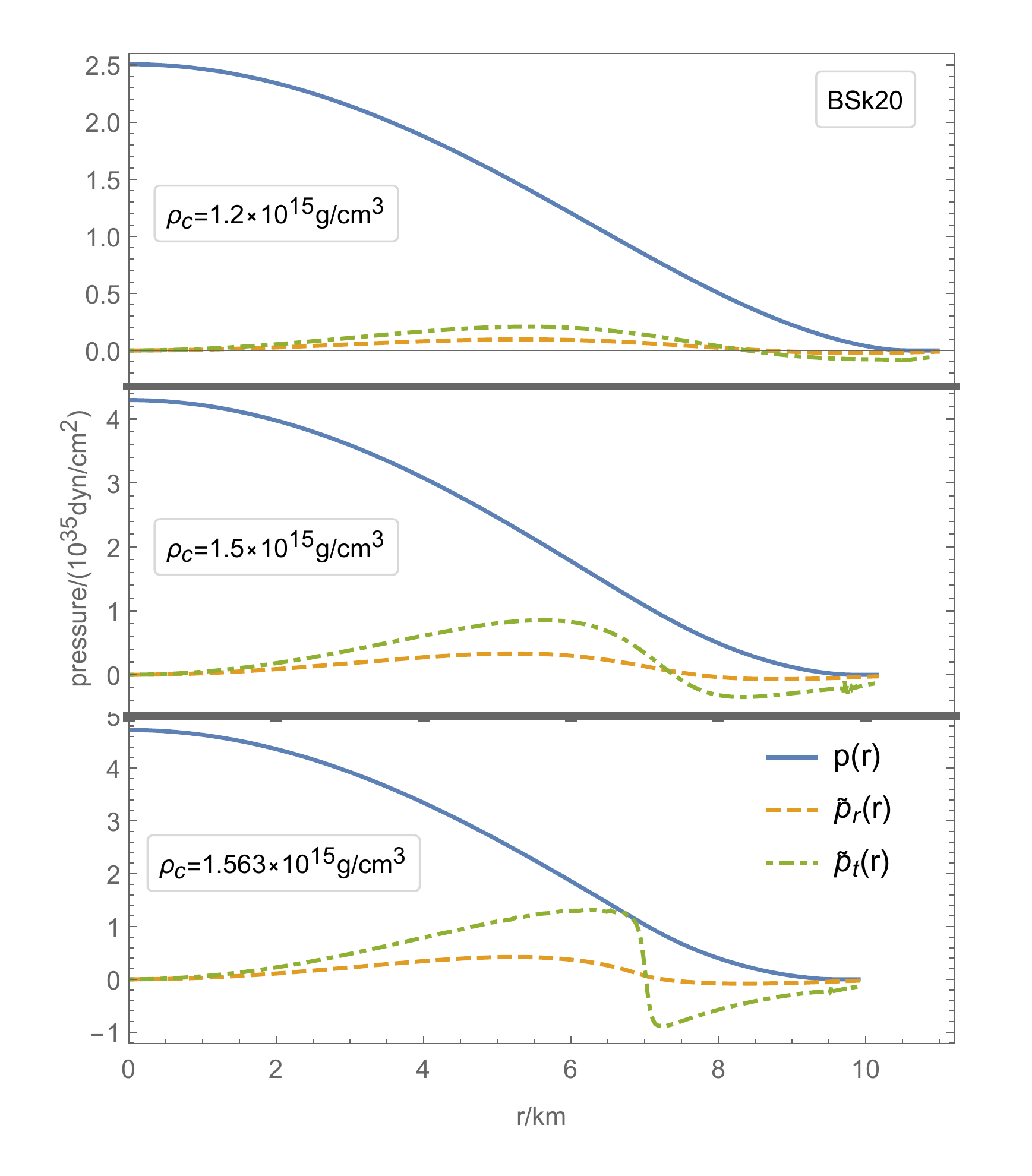}
    \end{subfigure}
    \begin{subfigure}{0.49\linewidth}
        \includegraphics[width=\linewidth]{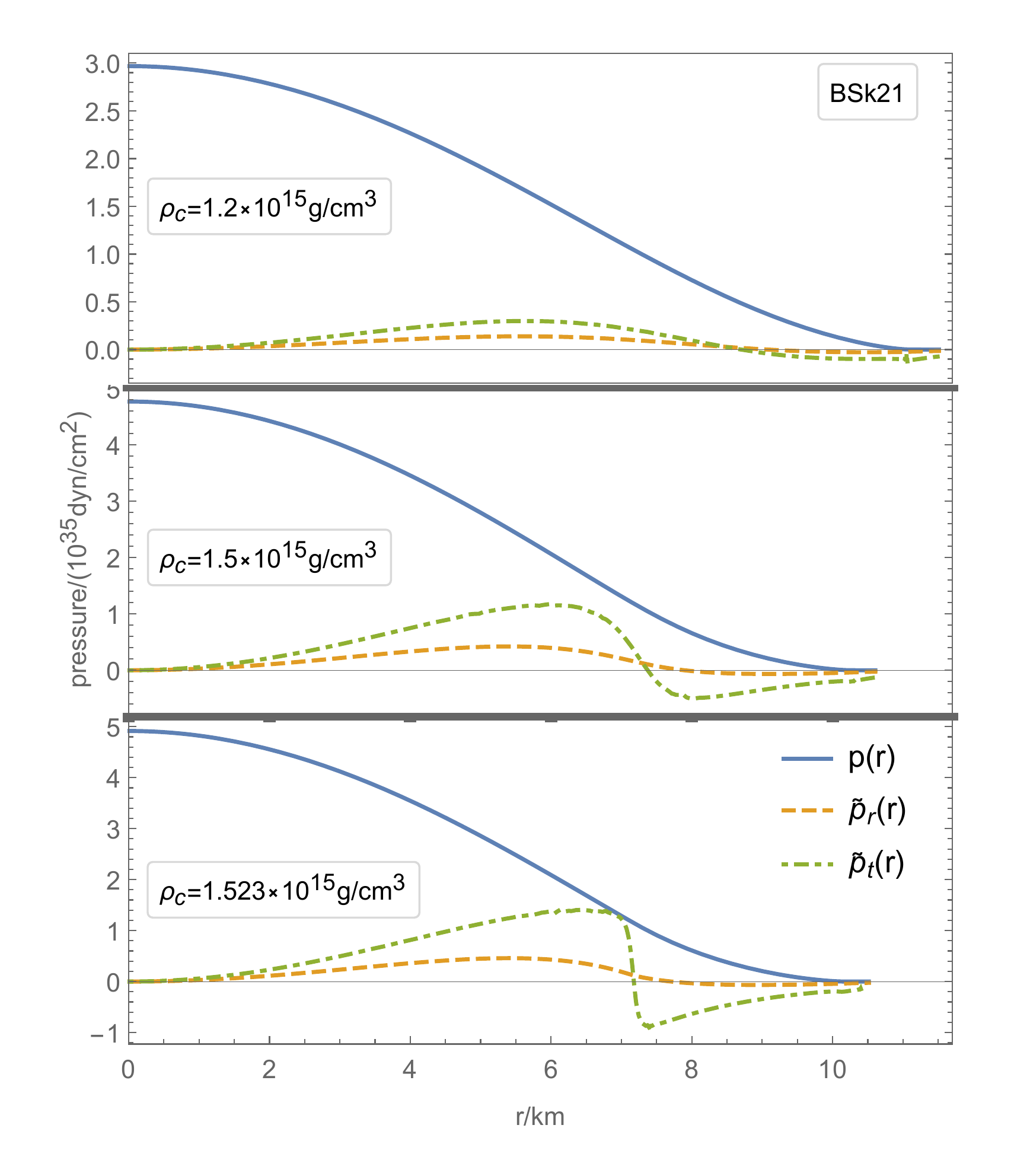}
    \end{subfigure}
    \caption{Pressure profiles, as well as the profiles of the effective radial and transverse pressures of the $f(T)$ \textit{fluid}, for SLy, BSk19, BSk20, and BSk21 EOSs
        with $\alpha=10r_g^2$ and several representative values of $\rho_c$.}
    \label{effp_plus}
\end{figure*}

For a given central density $\rho_c$,
we integrate the system of Eqs. \eqref{eomsph} and \eqref{TOV} using the EOSs described in the previous subsection.
When $\alpha>0$, there seems to be an upper limit for $\rho_c$ beyond which the numerical system cannot converge stably.
In Figs. \ref{effrho_plus} and \ref{effp_plus}, we show the density and pressure profiles,
as well as the profiles of the effective density $\tilde{\rho}$ and pressures $\tilde p_r,\tilde p_t$ of the $f(T)$ \textit{fluid} given in Eq. \eqref{effrhop},
for the aforementioned EOSs with $\alpha=10r_g^2$ and some representative values of $\rho_c$, where $r_g=G M_\odot/c^2\simeq 1.48\times10^5\:\text{cm}$.

One can see that for all the cases the density $\rho$ and pressure $p$ decrease outwardly and reach zero at some finite radius,
defining naturally the surface radius of the star.
Hence the stellar matter is indeed confined in a finite region independently of EOS.
Moreover, by comparing with Fig. \ref{gamma},
one can see that there will be a stable core where the adiabatic index $\Gamma>4/3$,
as long as the central density $\rho_c$ is set high enough ($\gtrsim 10^{15}\mathrm{g/cm^3}$).
And since the density will continuously reach zero near the surface,
there will be a crust where $\Gamma<4/3$.

On the other hand,
the effective density $\tilde{\rho}$, radial and transverse pressures $\tilde p_r,\tilde p_t$ of the $f(T)$ \textit{fluid}
vanish at the center of the star
and increase outwardly for the inner part of the stellar interior.
They reach their maxima at some point inside the star and start to decrease and change sign within the radius of the star.
At the surface of the star, while the effective radial pressure $\tilde p_r$ returns to zero,
the effective density $\tilde{\rho}$ and transverse pressure $\tilde{p}_t$ remain negative,
indicating a likely different external solution for nonlinear $f(T)$ gravity than the Schwarzschild vacuum in GR.
When $\alpha=0$,
$\tilde{\rho}$, $\tilde p_r$, and $\tilde p_t$ always vanish as mentioned before.
In this case, the solution reduces to that of GR and the external spacetime reduces to the Schwarzschild vacuum.

The last panel of each graph in Figs. \ref{effrho_plus} and \ref{effp_plus} corresponds to the value of $\rho_c$ close to the stably integrable limit of the numerical system,
which, for $\alpha=10r_g^2$, are $1.678,\: 1.859,\: 1.563$, and $1.523$ (in the unit of $10^{15}\mathrm{g/cm^3}$)
for SLy, BSk19, BSk20, and BSk21 EOSs, respectively.
The profiles of the effective $f(T)$ \textit{fluid} in these critical cases are qualitatively similar
in that $\tilde{\rho},\tilde{p}_t$ change abruptly at some point,
indicating rigid systems that lead to the breakdowns of the numerical procedures.
The core region of the NS defined by the steplike phase transition of the effective $f(T)$ \textit{fluid}
is then formed by a total effective fluid that has a slightly increasing density $\rho + \tilde{\rho}$.
This behavior of the mixture material of neutron matter and the effective $f(T)$ \textit{fluid}
is similar to that in the polytropic model where it is referred to as mimicking of an incompressible matter \cite{PhysRevD.98.064047}.

From Fig.\ref{effp_plus} or more explicitly, from calculation, one can easily find that
the central pressures for different EOSs are similar, i.e., $\sim 4.8\times 10^{35}~\text{dyn}/\text{cm}^2$ for the critical cases.
This can be interpreted as follows.
It is the pressure that balances the gravitation
which, in the current cases, is given by the same $f(T)$ model.
If the numerical system describing this gravitation has any sharp transition,
its occurrence will be most likely dominated by the initial value of pressure rather than density.
This correlation between pressure and the behavior of the numerical system
may be extended to the noncritical cases.
However, it does not mean similarity among the NS structures with different EOSs.
On the contrary, softer stellar matter will be compressed more tightly by the same level of gravitation
and hence the NS will have smaller radius.
From Figs. \ref{effrho_plus} and \ref{effp_plus},
one can see that for BSk19, SLy, BSk20, and BSk21 EOSs, in turn, describing stellar matter from soft to stiff,
the radii of NSs generally vary from relatively smaller to larger.

Although all models with positive $\alpha$'s have qualitatively similar behaviors and we have chosen a representative value $\alpha=10r_g^2$ to depict in Figs. \ref{effrho_plus} and \ref{effp_plus},
quantitatively, different values of $\alpha$ will no doubt affect the behaviors of the numerical systems.
For example, when $\alpha=5r_g^2$, the stably integrable limits of $\rho_c$ will be raised to $2.474,\: 2.670,\: 2.277$, and $2.329$
(in the unit of $10^{15}\mathrm{g/cm^3}$) for SLy, BSk19, BSk20, and BSk21 EOSs, respectively.
And when $\alpha=r_g^2$, the corresponding limit values are $8.628,\: 8.384,\: 7.935$, and $8.239$ (in the unit of $10^{15}\mathrm{g/cm^3}$), respectively,
which are all beyond the causal limits of the EOSs.
It is obvious that this breakdown point is independent on the causal limit.
For smaller $\alpha$, e.g., $\alpha=r_g^2$, one can choose any central density $\rho_c$ that respects causality,
while for greater $\alpha$, e.g., $\alpha=10r_g^2$, $\rho_c$ is confined by the stability and integrability of the system.
Moreover, the stably integrable limit of $\rho_c$ seems to be higher for a smaller modification.
We check $f(T)=T\e^{\alpha T}$ model for confirmation of this observation.
Since the torsion scalar is generally negative for the gravitational system under consideration,
the factor $\e^{\alpha T}$ leads to a smaller modification than the term $\alpha T^2$ for the same value of $\alpha$.
The limits of $\rho_c$ for the steplike behavior of the $f(T)$ \textit{fluid} are raised in $T\e^{\alpha T}$ model as expected.

\begin{figure*}[h!]
    \centering
    \begin{subfigure}{0.49\linewidth}
        \includegraphics[width=\linewidth]{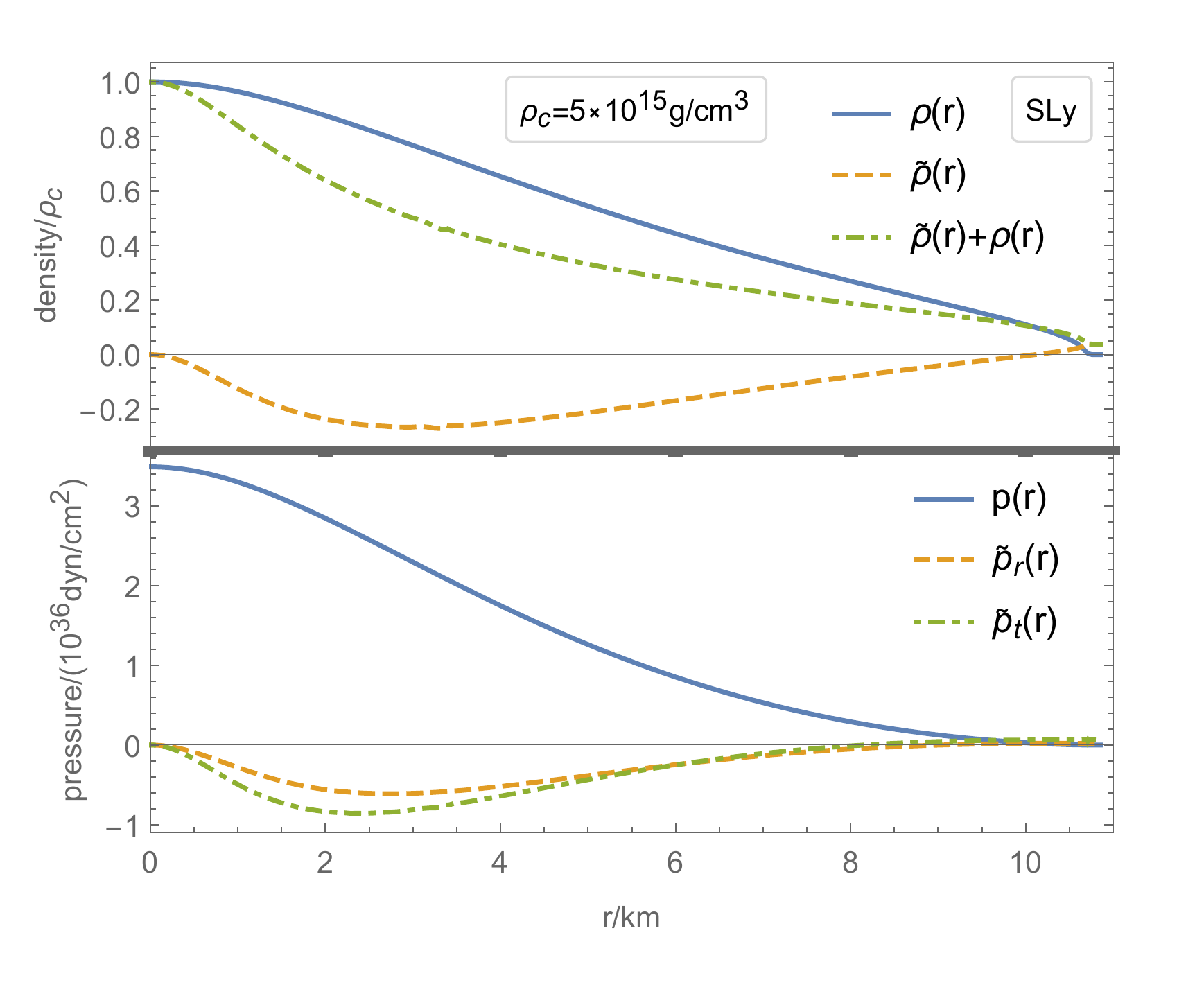}
    \end{subfigure}
    \begin{subfigure}{0.49\linewidth}
        \includegraphics[width=\linewidth]{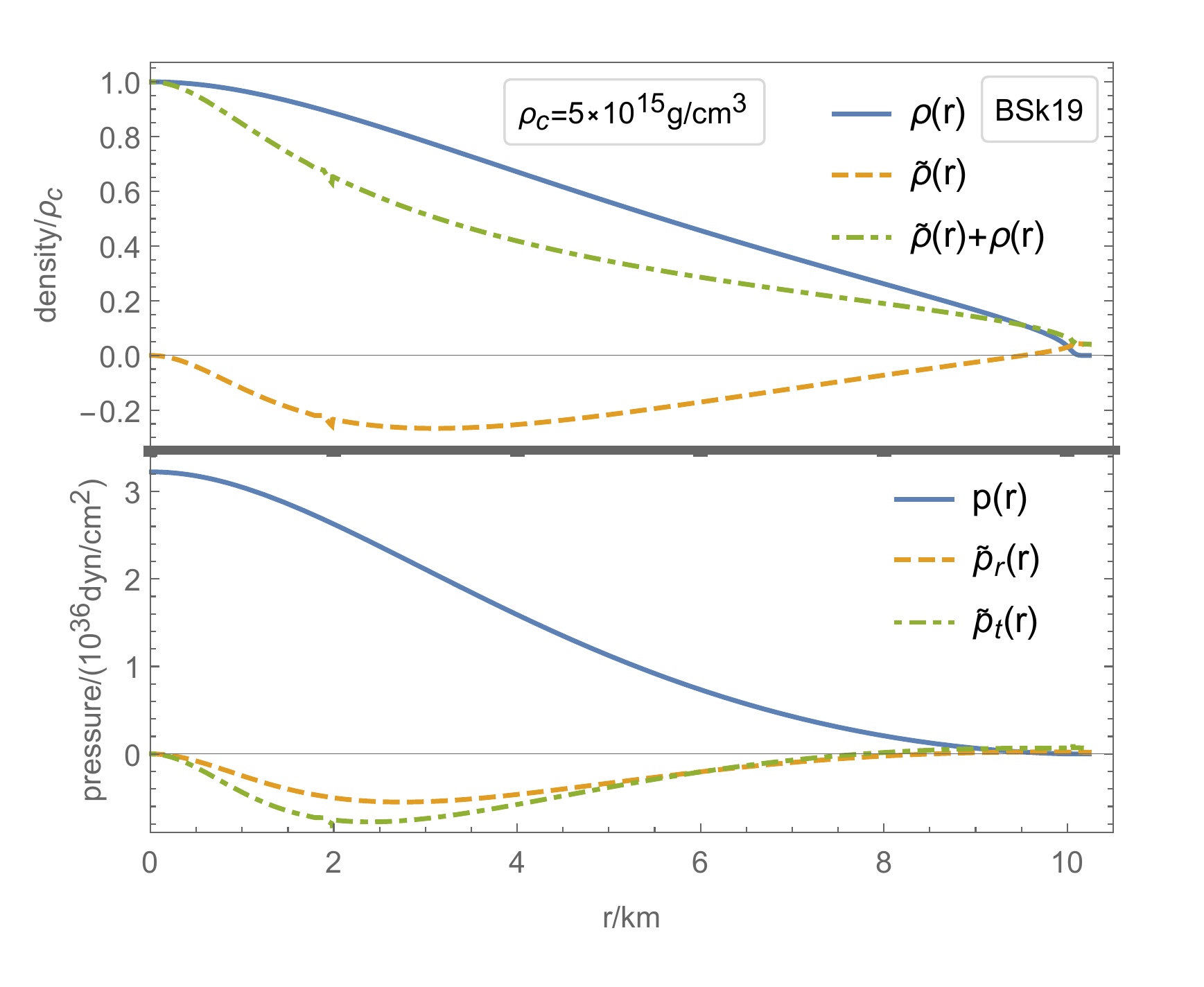}
    \end{subfigure}
    \begin{subfigure}{0.49\linewidth}
        \includegraphics[width=\linewidth]{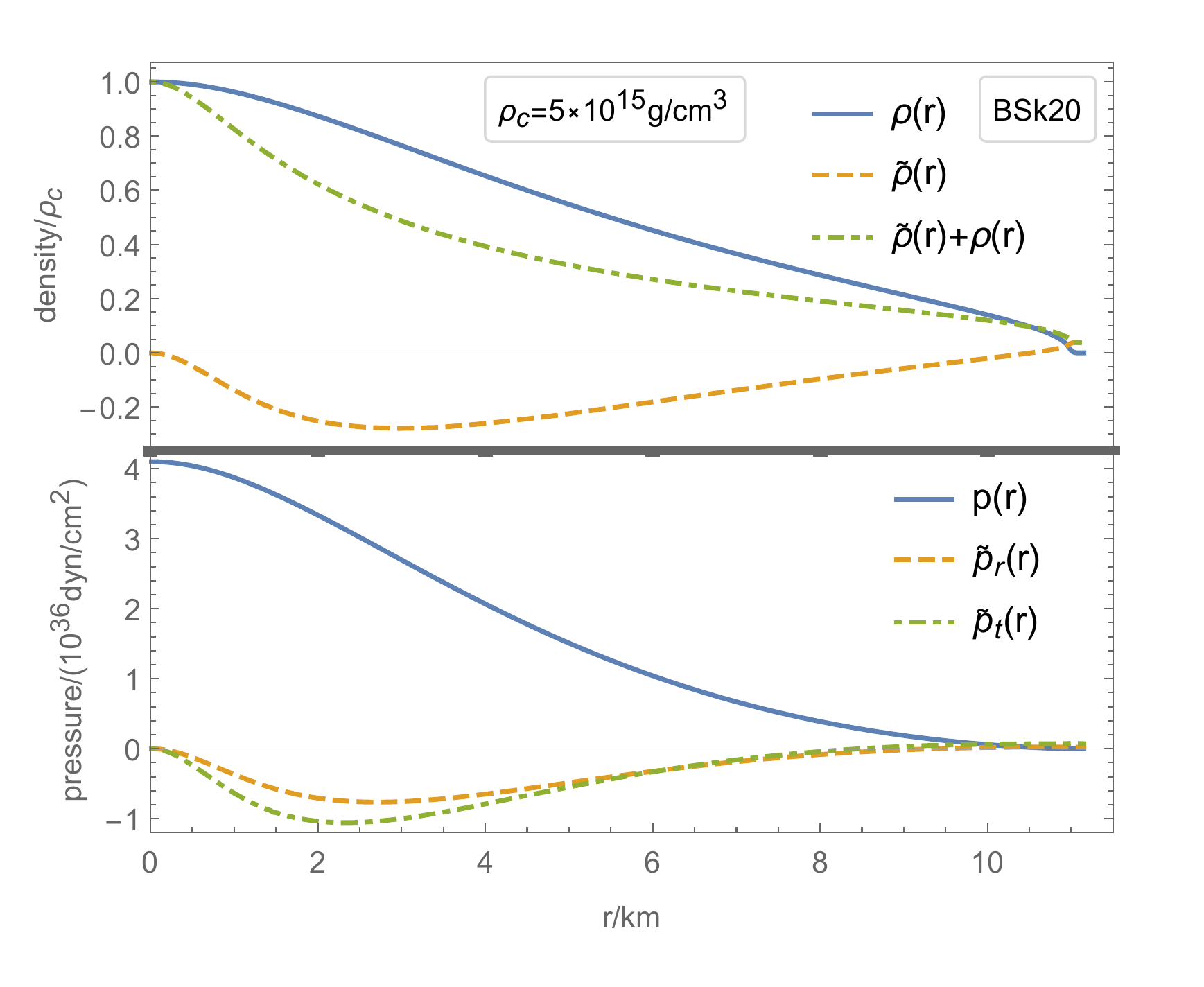}
    \end{subfigure}
    \begin{subfigure}{0.49\linewidth}
        \includegraphics[width=\linewidth]{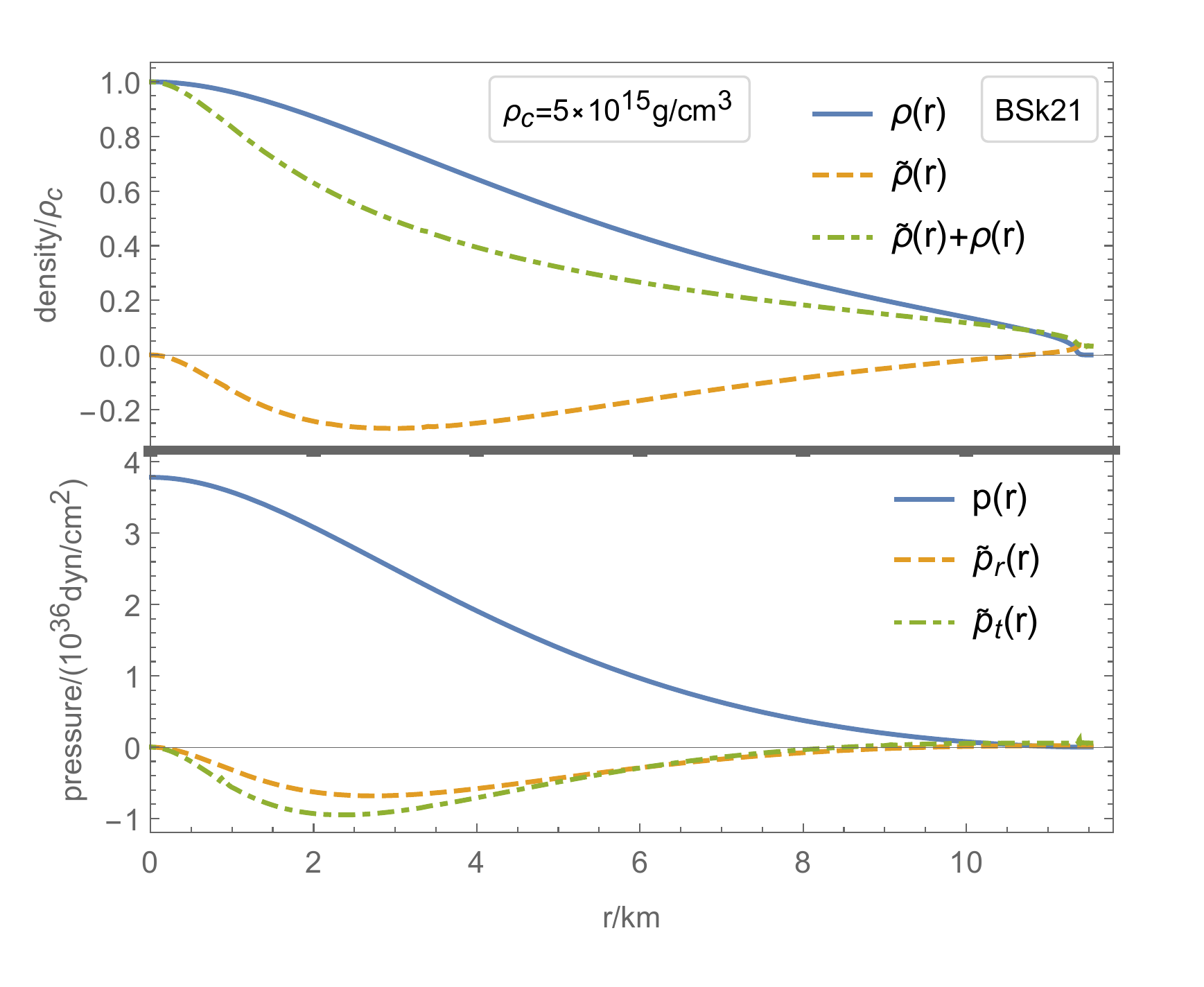}
    \end{subfigure}
    \caption{Density and pressure profiles for SLy, BSk19, BSk20, and BSk21 EOSs
        with $\alpha=-10r_g^2$ and $\rho_c=5.0\times 10^{15}\mathrm{g/cm^3}$.
        The radii of the NSs are $10.86\text{km},\: 10.24\text{km},\: 11.15\text{km}$, and $11.53\text{km}$, respectively.}
    \label{effrho_minus}
\end{figure*}

In Fig. \ref{effrho_minus}, we present the density and pressure profiles,
as well as the profiles of the effective density and pressures of the $f(T)$ \textit{fluid},
for the EOSs of SLy and BSk family with $\alpha=-10r_g^2$ and $\rho_c=5.0\times 10^{15}\text{g/cm}^3$ as a representative case of negative $\alpha$.
The matter density and pressure also decrease outwardly and vanish at a finite surface radius of the star.
The profiles for different EOSs with a negative $\alpha$ are qualitatively similar to each other,
and also to the polytropic model \cite{PhysRevD.98.064047}.
The effective density and pressures of the $f(T)$ \textit{fluid}, in contrast with the cases of positive $\alpha$,
are negative for most part of the interior region of the star.
They change sign near the surface and remain positive at the radius of the star,
which, like the cases of positive $\alpha$, indicates a possibly different external solution of $f(T)$ gravity than the Schwarzschild one.
No steplike changes can be seen in profiles of the cases of negative $\alpha$.
More precisely, the sharp transition of $f(T)$ \textit{fluid} only happens when the modification term is positive.
This can be checked in the $f(T)=T+\alpha T^3$ model,
where the steplike behavior occurs when $\alpha<0$.

\section{Masses and radii}
\label{mrs}
\subsection{Mass-radius relation}
The term \textit{stellar mass} generally means the total mass (or energy) of the stellar gravitation system
that may be measured at a distance.
Due to the non-linearity of the gravitation field, this total (or effective, active) mass is
usually a combination of the material (or passive, rest) mass and the energy of the corresponding gravitation field.
In spherical symmetry with a metric written in Eq.\eqref{sphmetric},
the material mass within a hypersurface $\Sigma$ of radius $r$ is given by
\begin{equation}
    \label{massint}
    m(r)=\int_\Sigma\star\rho=4\pi\int_0^r\rho(x)\e^{B(x)}x^2\diff x,
\end{equation}
where $\star 1$ is the volume form and $\rho$ is the energy density.
The definition of the active mass, however, relies on the gravitation theory.
In GR, the active mass of an asymptotically flat spacetime can be defined at spatial infinity by
the Arnowitt-Deser-Misner (ADM) mass \cite{Arnowitt:1959ah}
and at null infinity by the Bondi-Sachs (BS) mass \cite{Bondi:1962px,Sachs:1962wk}.
Within a finite hypersurface $\Sigma$, the Misner-Sharp (MS) mass \cite{Misner:1964je}
gives a definition of the active mass that can reduce to the ADM or BS mass asymptotically at the corresponding infinity \cite{Hayward:1994bu}.
For a spherically symmetric spacetime either dynamic or static,
the metric can be written in the following coordinates,
\begin{equation}
    \label{sphmetric2}
    \diff s^2=I_{ab}\diff x^a\diff x^b+r^2\diff\Omega_2^2,
\end{equation}
where $\diff\Omega_2^2$ represents a unit 2-sphere,
$I_{ab}$ is the induced metric that does not depend on the inner coordinates of the unit 2-sphere,
and $a,b,\cdots$ run from 0 to 1.
The MS mass in GR then can be written as
\begin{equation}
    \label{MSGR}
    M_\text{MS}^{(\text{GR})}=\frac{r}{2G}\left( 1-I^{ab}\partial_ar\partial_br \right).
\end{equation}
One can see that Eq.\eqref{MSGR} does not involve explicitly any term of material source, e.g., energy density $\rho$ or pressure $p$.
In fact, the MS mass, as well as the ADM mass and BS mass, has utilized the gravitational field equations
to transform the combination of the material mass and the gravitational energy into the geometric terms of the spacetime.
A known solution of the metric and the material source can, of course, reproduce the active mass written in terms of $\rho$ and $p$.
But they depend closely on the gravitation theory, or more precisely, the field equations
since the active mass is in fact the solution to the field equation in a different, integral form.

Therefore, in modified gravities with different field equations than the Einstein equation,
the definition of the active mass, or, the MS mass in a finite spacetime with spherical symmetry,
needs to be reconsidered,
which is generally thought to be much easier and more definitive when an explicit solution is known \cite{PhysRevD.98.064047}.
In $f(R)$ gravity, this issue is studied in Refs. \cite{Cai:2009qf,Zhang:2014goa}.
Following the idea of these works, in $f(T)$ gravity, one starts from the physical meaning of the active mass,
i.e., the unified first law \cite{Hayward:1997jp},
\begin{equation}
    \label{massms}
    \diff M_\text{MS}=\mathcal A \psi_a\diff x^a+W\diff\mathcal V,
\end{equation}
where $\mathcal A$ and $\mathcal V$ are the surface area and volume of the space region being considered, respectively.
$W$ is the work term and $\psi_a$ is the energy supply term, which are given by
\begin{equation}
    \label{work}
    W=-\frac12I^{ab}\mathcal T_{ab},\quad \psi_a=\mathcal T_a^b\partial_b r+W\partial_ar.
\end{equation}
Note that although Eq.\eqref{work} does not involve any explicit feature of the gravitation theory,
they are consistent with the Einstein equation.
If Eq.\eqref{work} is assumed to still hold in modified gravities \cite{Cai:2009qf,Zhang:2014goa},
one then immediately obtains,
\begin{equation}
    \label{massms1}
    \diff M_\text{MS}=\mathcal A I^{ab}\left( \mathcal T_{ab}\partial_cr-\mathcal T_{ac}\partial_br \right)\diff x^c.
\end{equation}
In the static case with the usual spherical coordinates,
Eq.\eqref{massms1} gives the intuitive definition of the active mass
\begin{equation}
    \label{massms2}
    \diff M_\text{MS}=\mathcal A\mathcal T_0^0\diff r=4\pi r^2\rho\diff r.
\end{equation}
Further derivation of the MS mass involves substitution of $\mathcal T_0^0$ in Eq.\eqref{massms2}
with the field equation, i.e., Eq.\eqref{fieldeq} in the current case.
This, however, is effectively the procedure of solving the field equation for external solutions
and is beyond the scope of this work.
Nonetheless, since we have already obtained the energy density profile $\rho(r)$ numerically,
Eq.\eqref{massms2} is sufficient to define the active mass of an NS in $f(T)$ gravity.
Due to the fact that $\rho(r)$ vanishes at the stellar surface $\mathcal R$,
the active stellar mass is given by
\begin{equation}
    \label{massns}
    M=4\pi\int_0^\mathcal R\rho(r)r^2\diff r.
\end{equation}
We note here that although we have not presumed any known external solution,
the matching condition at the surface should be respected.
As mentioned before, if one continues to substitute $\mathcal T_0^0$ in Eq. \eqref{massms2} with the field equation \eqref{fieldeq}
and carries out the integral,
a solution of metric in terms of $M_\text{MS}$ can be, in principle, obtained.
And as long as the field equation is continuous at $\mathcal T_0^0=0$,
this solution of metric will be smooth at the surface.
According to the profiles of the effective $f(T)$ \textit{fluid} presented in the previous section,
the external field equation will become $G_{\mu\nu}=\tilde{\mathcal T}_{\mu\nu}$ with nonzero $\tilde{\mathcal T}_{\mu\nu}$.
Hence the spherical vacuum may not be the same as the Schwarzschild one.
Some studies on the spherical vacuum in $f(T)$ gravity can be found in Refs. \cite{PhysRevD.94.124025,Golovnev:2021htv,Pfeifer:2021njm,DeBenedictis:2022sja}.

\begin{figure*}[h!]
    \centering
    \begin{subfigure}{0.49\linewidth}
        \includegraphics[width=\linewidth]{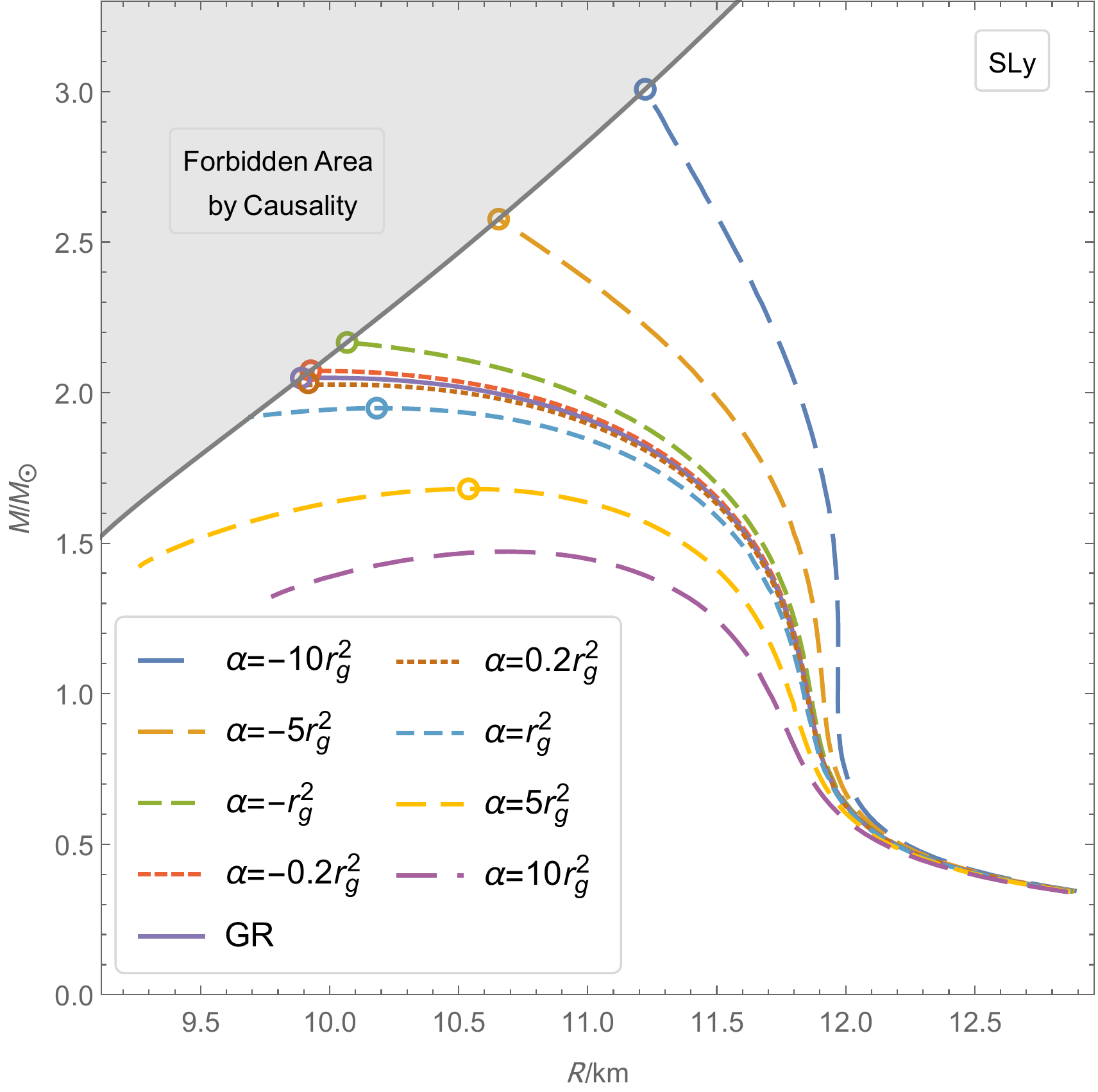}
    \end{subfigure}
    \begin{subfigure}{0.49\linewidth}
        \includegraphics[width=\linewidth]{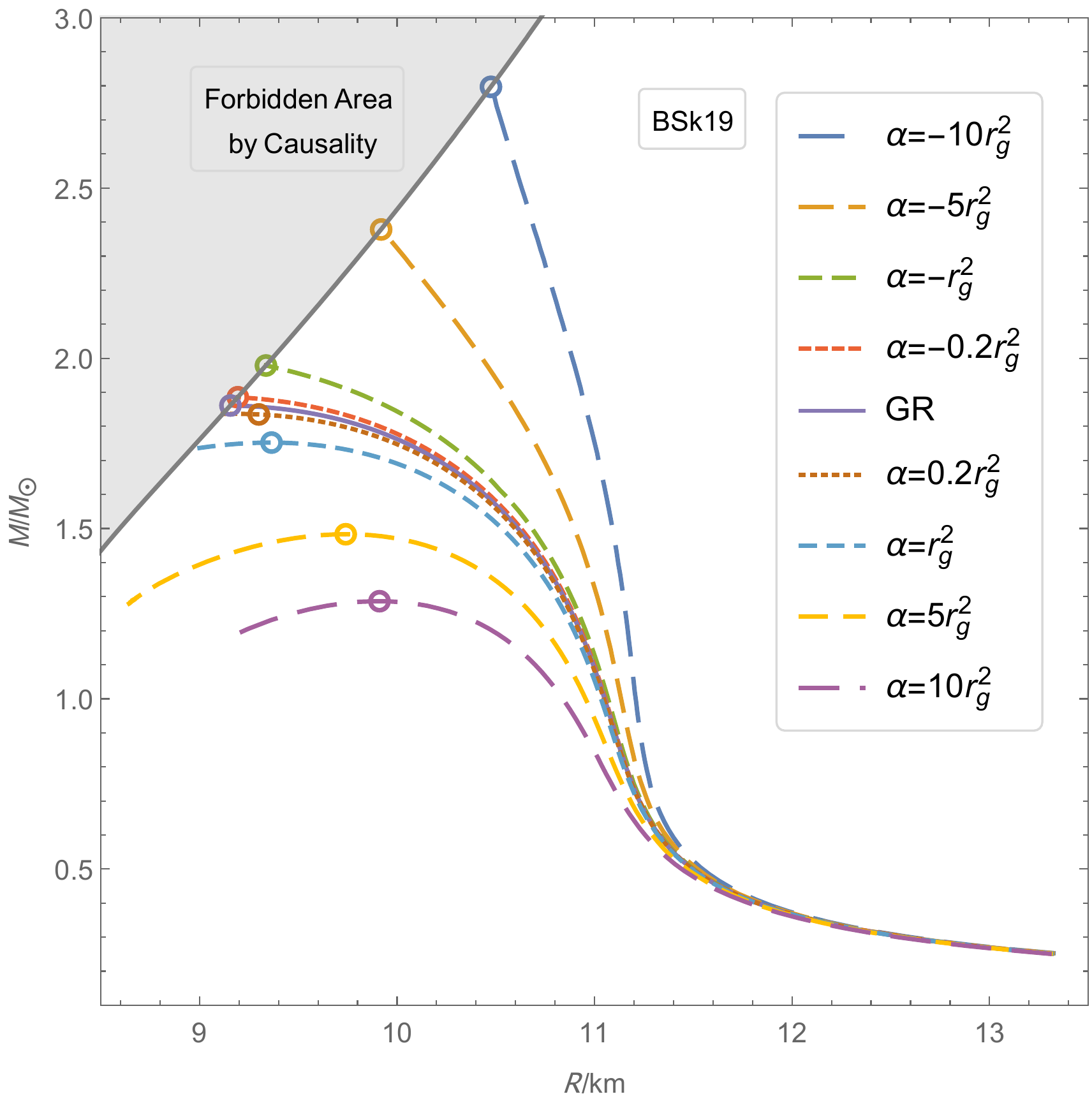}
    \end{subfigure}
    \begin{subfigure}{0.49\linewidth}
        \includegraphics[width=\linewidth]{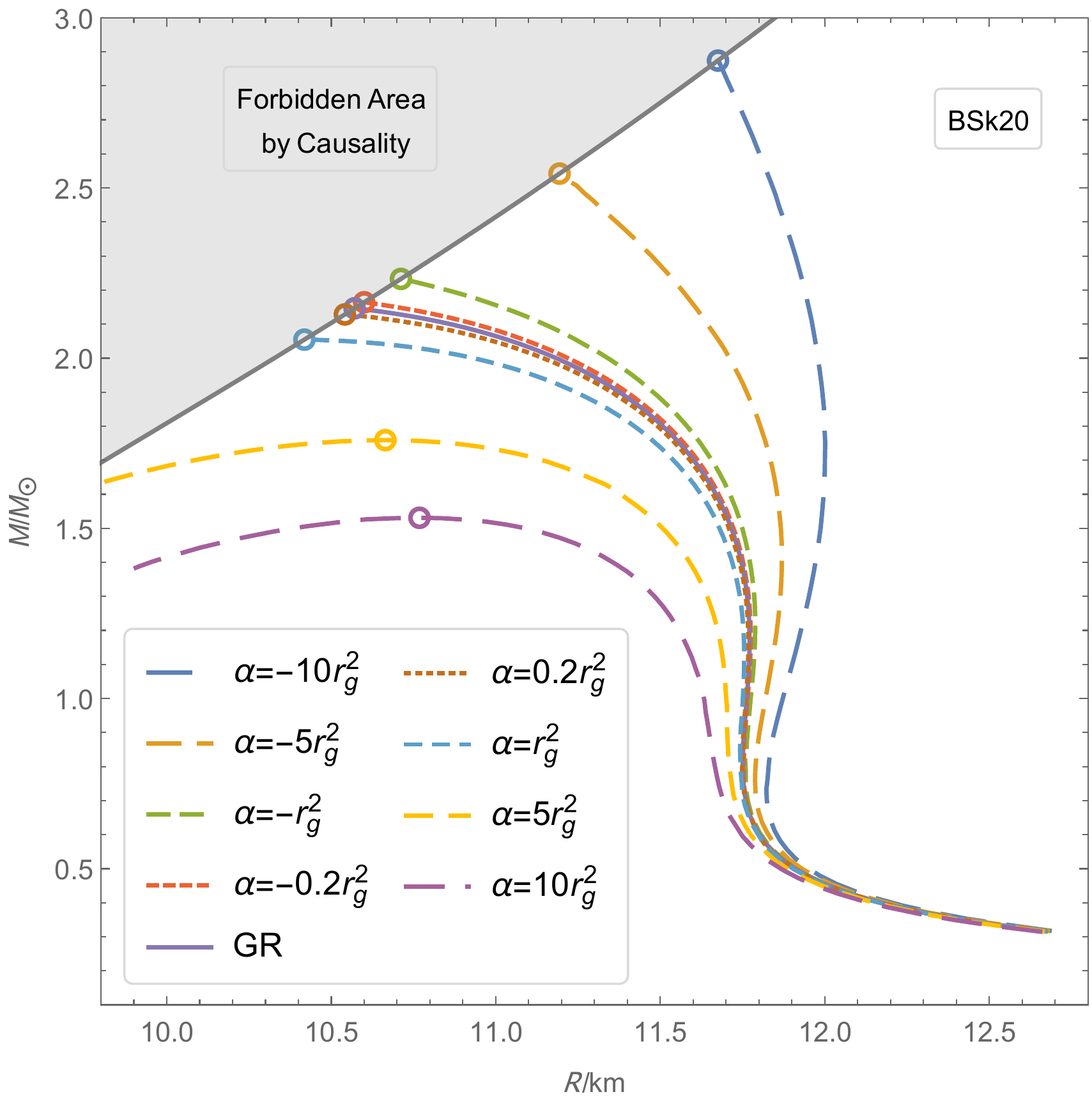}
    \end{subfigure}
    \begin{subfigure}{0.49\linewidth}
        \includegraphics[width=\linewidth]{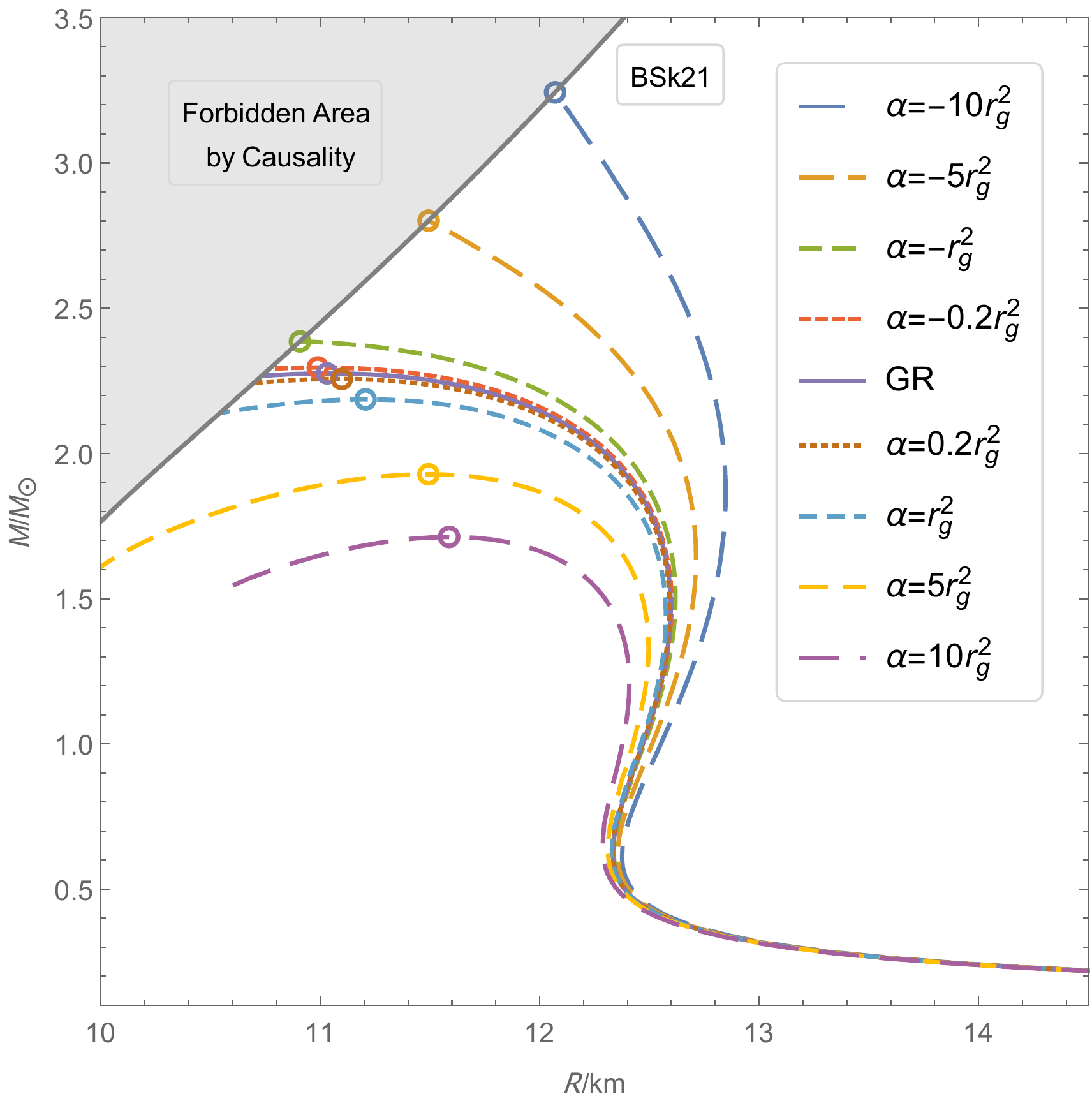}
    \end{subfigure}
    \caption{Mass-radius curves for SLy, BSk19, BSk20, and BSk21 EOSs,
        considering various representative values of $\alpha$.
        The curve corresponding to GR is recovered when $\alpha=0$.}
    \label{figMR}
\end{figure*}

We now proceed with Eq. \eqref{massns} and calculate the NS masses.
For a given central density $\rho_c$ and hence a numerically determined profile $\rho(r)$,
a pair $\{M,\mathcal R\}$ can be found.
Thus, one obtains an $M-\mathcal R$ curve on which every point corresponds to a different $\rho_c$.
In Fig. \ref{figMR} we present the $M-\mathcal R$ curves for some representative values of $\alpha$
and all the EOSs described in the previous subsection.
At relatively low central density around or below $\rho_c\sim10^{13}\text{g/cm}^3$,
the radius and mass only show subtle differences for various values of $\alpha$ with a given EOS.
This may indicate that a unified treatment, possibly a nonrelativistic one like Newtonian star,
is sufficient for the models at this order of density.
However, stars at or below this density may not be described by these EOSs considered
in that the adiabatic indexes may fall below the stability line $4/3$ as discussed in the previous section.
At neutron stellar density around or above $\rho_c\sim10^{15}\text{g/cm}^3$,
changes of the active mass and radius of the star can be observed.
Since there are upper limits of the densities due to the causality condition,
the numerical integration cannot go to arbitrary high central density $\rho_c$.
This leads to forbidden areas in the mass-radii plots.
The NS in $f(T)$ gravity with a central density respecting the causality condition
will not have the mass-radius pair in this area.

For positive $\alpha$'s, the numerical procedure continues till the system reaches the causal limit
or the stably converging limit as discussed in the previous subsection.
Relatively small $\alpha$ (see the $\alpha=0.2 r_g^2$ lines in each panel) results in
$M-\mathcal{R}$ curves that are quite close to GR cases that can also been found in, e.g., Refs. \cite{Douchin:2001sv,Haensel:2004nu,Potekhin:2013qqa},
somewhat validating the stability of the system and the reliability of the numerical procedure.
As $\alpha$ grows and hence the model moves away from GR, significantly less stellar matter can be contained in a given radius.
In these cases, there exist such critical configurations that the stellar masses reach their maxima.
If the ratio between pressure and energy density goes beyond this critical point, the structure of the star may become unstable.
The critical configuration of different values of $\alpha$ for each EOS, if exists, are marked in Fig. \ref{figMR} with circle symbols.
The detail values of the critical configurations can be found in \ref{appcri}.

For negative $\alpha$'s, the numerical system allows the central densities reach the causal limit.
Relatively small $|\alpha|$ (see the $\alpha=-0.2 r_g^2$ lines in each panel)
may still present a peak in $M-\mathcal R$ curve as GR.
But as the central densities are set to be higher,
the upper bounds of the stellar mass are determined by the causal limits of the central densities $\rho_c$.
As examples, in the cases of $\alpha=-10r_g^2$ (the blue dashed line in each panel of Fig. \ref{figMR}),
the upper bounds of the stellar mass are $3.008M_\odot$, $2.797M_\odot$, $2.874M_\odot$, and $3.243M_\odot$
for SLy, BSk19, BSk20, and BSk21 EOS, respectively.
As $\alpha$ decreases, more stellar matter can be contained in a given radius,
resulting in a more compact star.
A qualitatively analogous pattern is reported for the material mass of the polytropic model of stars in $f(T)$ gravity \cite{PhysRevD.98.064047}.

\subsection{Observation constraints}
\begin{figure*}[h!]
    \centering
    \begin{subfigure}{0.49\linewidth}
        \includegraphics[width=\linewidth]{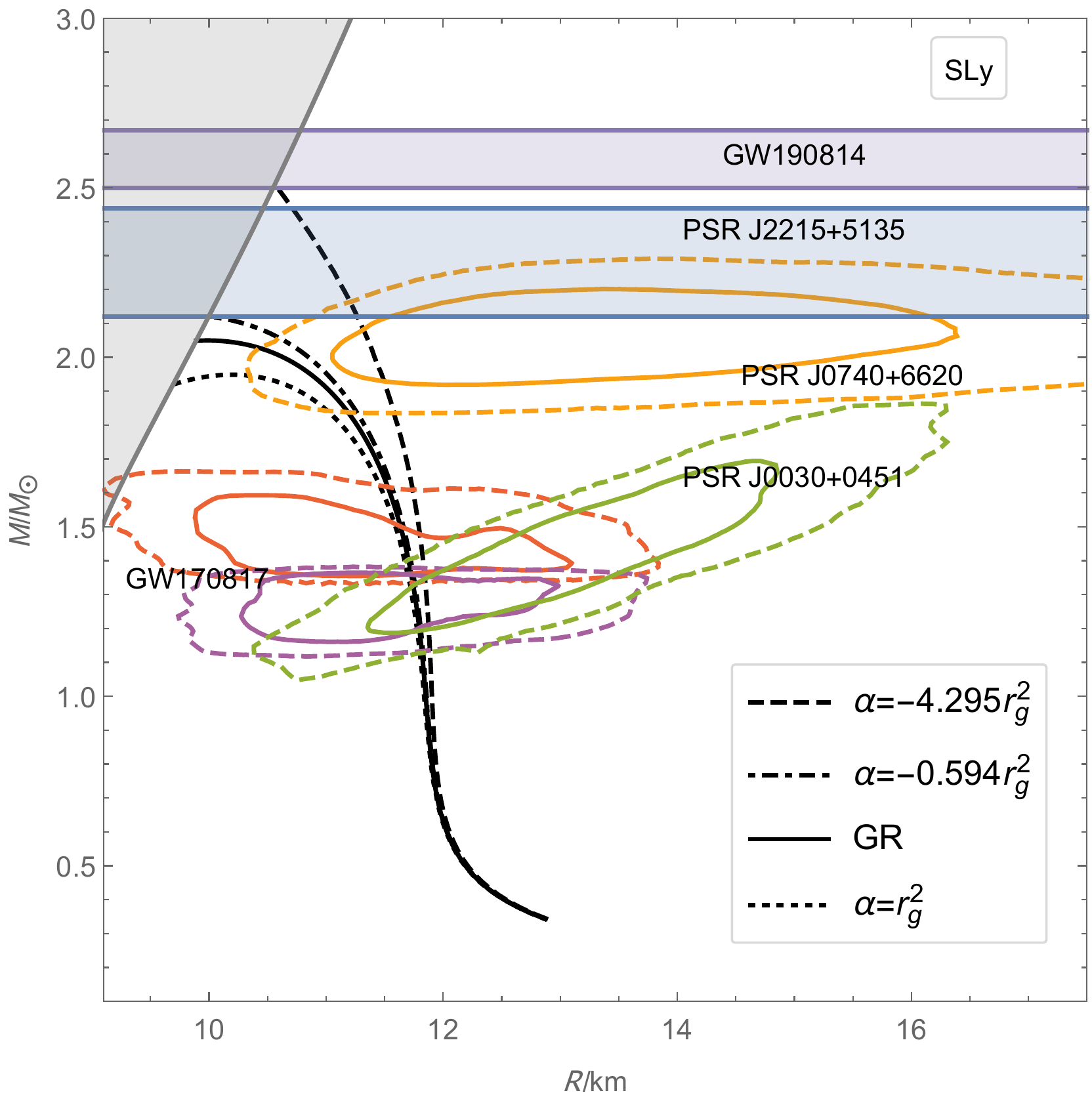}
    \end{subfigure}
    \begin{subfigure}{0.49\linewidth}
        \includegraphics[width=\linewidth]{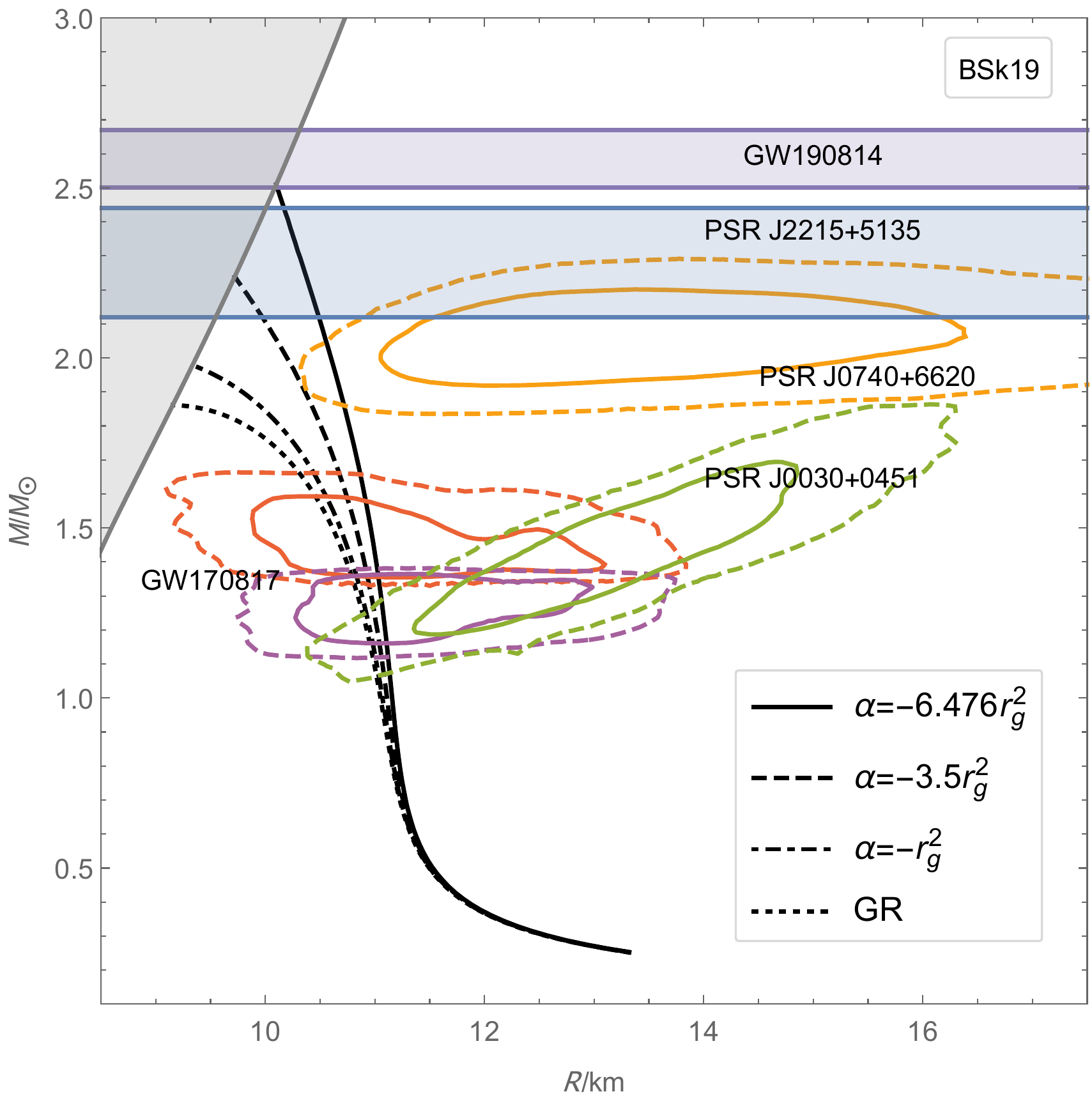}
    \end{subfigure}
    \begin{subfigure}{0.49\linewidth}
        \includegraphics[width=\linewidth]{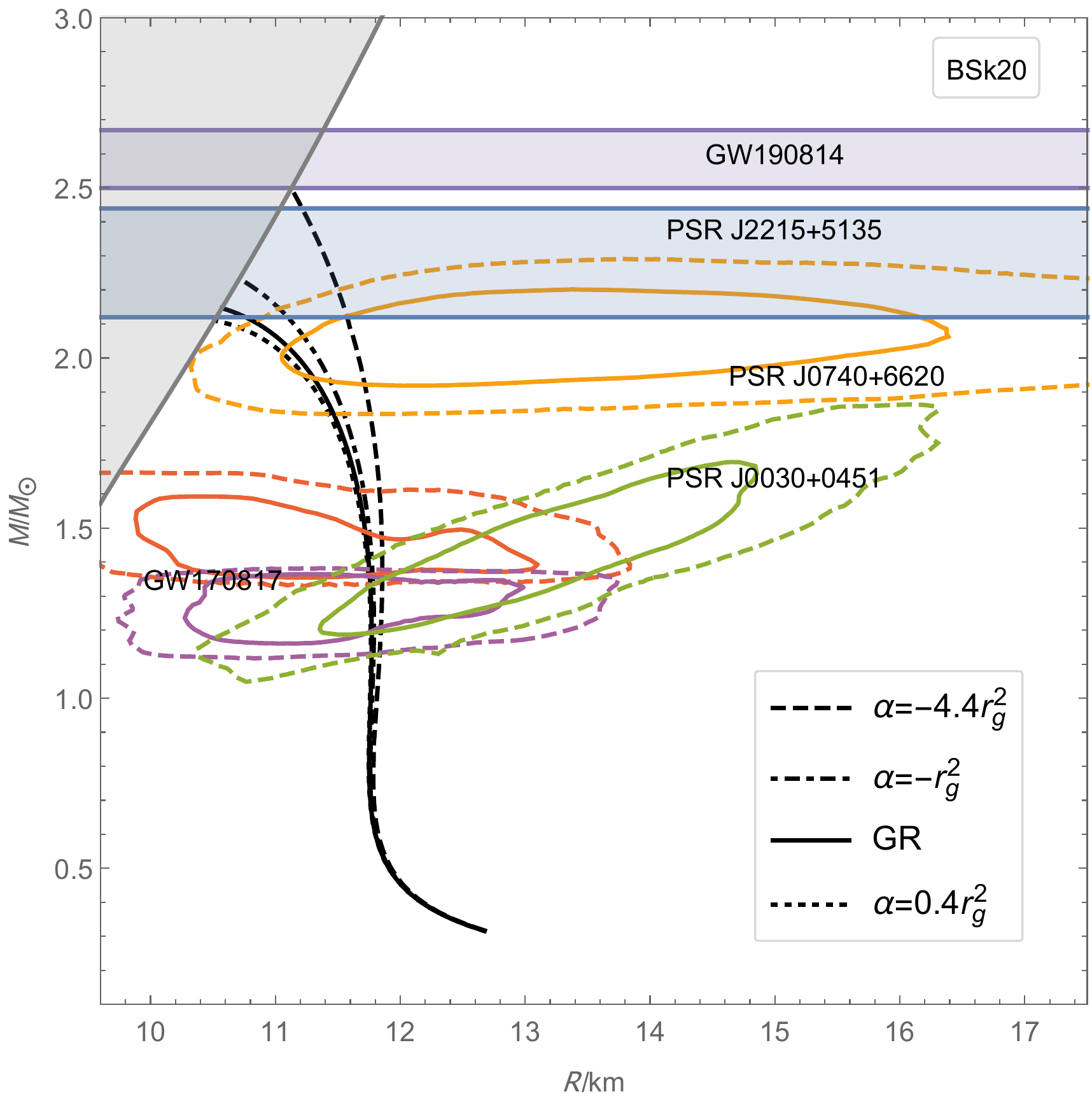}
    \end{subfigure}
    \begin{subfigure}{0.49\linewidth}
        \includegraphics[width=\linewidth]{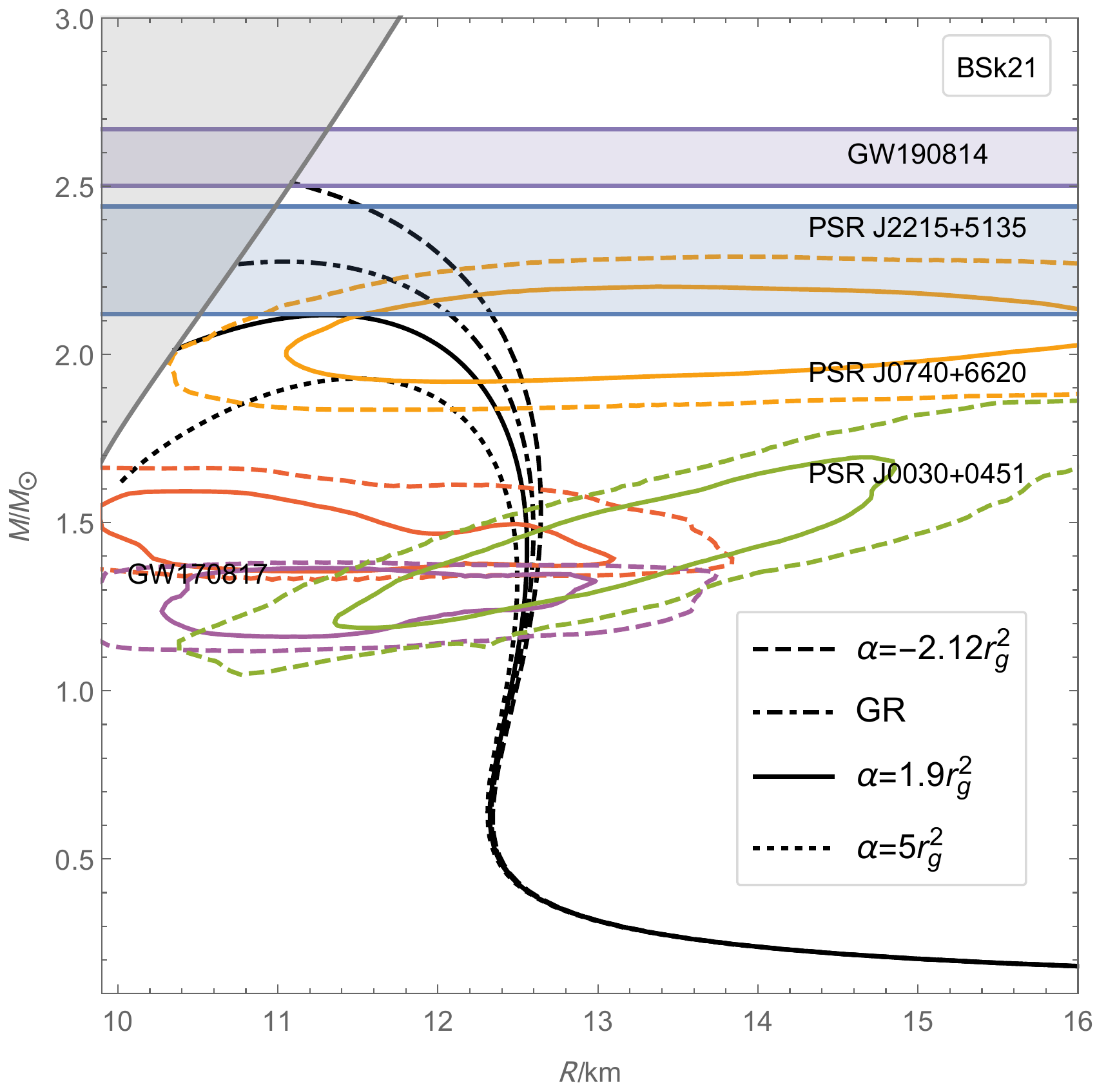}
    \end{subfigure}
    \caption{Mass-radius curves for SLy, BSk19, BSk20, and BSk21 EOSs,
        compared with the observation constraints.}
    \label{figcon}
\end{figure*}

In this subsection,
the mass-radius curves of the NSs in $f(T)$ gravity will be subjected to the joint constraint
from the observed massive pulsars and the gravitational wave events.
The detection of the first GW signal from a binary NS merger GW170817 by LIGO/Virgo collaboration \cite{LIGOScientific:2018cki},
together with its electromagnetic counterpart, GRB170817,
provides information of the masses and radii of the two NSs in this event.
Another GW event, GW190814 \cite{Abbott:2020khf}, indicates that a compact object, possibly an NS, with a mass of $2.59\pm0.08M_\odot$ may exist.
Besides this compact object, an NS with a mass of $2.27^{+0.17}_{-0.15}M_\odot$, hosted by PSR J2215+5135 \cite{Linares:2018ppq}, is reported as one of the most massive NSs known to date.
Moreover, joint observation of the mass and radius of the pulsar PSR J0030+0451 by NICER \cite{Miller:2019cac}
also provides independent constraint on NS properties.
By combining the data from NANOGrav 12.5-year dataset
with the orbital-phase-specific observations using the GBT,
another massive NS PSR J0740+6620 with a mass of $2.14^{+0.20}_{-0.18}M_\odot$ has been reported \cite{NANOGrav:2019jur}.
With the NICER and X-ray Multi-Mirror (XMM) X-ray observation, the radius of PSR J0740+6620 has also been measured \cite{Miller:2021qha}.
The constraints on the mass-radius relation from the aforementioned observations are presented in Fig.\ref{figcon}.
The inner solid contour and outer dashed contour for each set of observational data correspond to
the $1\sigma~(68.3\%)$ and $2\sigma~(95.4\%)$ confidential levels, respectively.
The GW170817 dataset produces two sets of contours, corresponding to two components of the binary.
Observation of PSR J2215+5135 and the GW event GW190814 only give constraints of the mass,
which are depicted as the blue and purple bands in the graphs, respectively.
We compare these constraints with the $M-\mathcal R$ curves of NSs in $f(T)$ gravity in Fig. \ref{figcon},
and obtain the constraints for the model parameter $\alpha$ for different EOSs.

For SLy, one can see that the GR curve cannot reach the mass range given by the dataset of PSR J2215+5135.
And the causal limit gives that when $\alpha\le-0.594 r_g^2$,
the curve can reach the mass range given by PSR J2215+5135.
If one considers the compact object in the GW event GW190814 to be an NS and wishes the mass range given by this dataset to be reached,
the model parameter $\alpha$ needs to be less than $-4.295 r_g^2$.
Therefore, for the NS model to accommodate the observation data,
the interception of constraints leads to the requirement that $\alpha<-4.295r_g^2$.

For BSk19, one can see that for the curve to pass the contours of PSR J0740+6620,
the model parameter $\alpha$ needs to be smaller than $-3.5r_g^2$.
In this case, the curve can also reach the mass range given by PSR J2215+5135.
Moreover, $\alpha<-6.476r_g^2$ is required for the causal limit of the curve reaches the mass range given by GW190814
and accommodate all the observation data.

For BSk20, the GR curve can already accommodate all the data except GW190814.
The models with $\alpha>0.4r_g^2$ predict that the NS mass
cannot reach the mass range given by the observation of PSR J2215+5135,
and hence are ruled out.
Moreover, if one considers the compact object in the GW event GW190814 to be an NS and wishes the mass range given by this dataset to be reached,
the model parameter $\alpha$ has to be less than $-4.4r_g^2$ so that
the upper bound of the NS model can reach the mass range given by GW190814.

For BSk21, once again the GR curve can already accommodate all the data except GW190814.
The requirement for the models to accommodate the observation of PSR J2215+5135 is $\alpha\le1.9r_g^2$ (see the solid line).
And an $\alpha$ smaller than $-2.12r_g^2$ can also enable the model to accommodate the GW190814 data for a possible NS.

\section{Conclusion and discussions}
\label{conclusion}
In this paper, we employ the realistic EOSs to investigate the NSs in $f(T)$ gravity.
In particular, we study the static and spherically symmetric configuration with neutron stellar matter described by SLy and BSk family of EOSs
in a simple but realistic nonlinear model $f(T)=T+\alpha T^2$.
This model is able to describe the inflationary universe that may has similarly extreme condition of the NS interior,
and hence is a reasonable example to consider NSs in $f(T)$ gravity.
For both positive and negative values of $\alpha$,
we show that the model indeed provides a compact star solution.
As depicted in Fig. \ref{figMR}, significant changes of stellar mass-radius relation can be seen
as $\alpha$ moves away from zero and hence the model departs from GR.
Moreover, regardless of the EOS, models with a negative $\alpha$ may support more matter for the compact star than in GR,
while less matter can be contained with a positive $\alpha$.
This may be understood qualitatively as follows.
In the interior region of an NS,
$f(T)$ model with a positive modification exerts a stronger gravity than GR for a given amount of matter.
Thus, at the same pressure level, e.g., degeneracy pressure of nuclear particles, less matter can be supported in such a case.
On the other hand, a negative modification may act as diminishment of the gravity,
or, if considered as an effective $f(T)$ \textit{fluid}, provide a pressure to resist the gravity;
hence, more matter can be supported.
This can also be seen in Figs. \ref{effp_plus} and \ref{effrho_minus}.
The effective pressures of the $f(T)$ \textit{fluid} are positive for most part of the interior of the star when $\alpha$ is positive,
while a negative value of $\alpha$ may lead to almost always negative effective pressures of the $f(T)$ \textit{fluid} inside the star.

In GR, due to the limit of nuclear degeneracy pressure, there exists a configuration that corresponds to a maximum stellar mass \cite{Rhoades:1974fn}.
In $f(T)$ gravity, this limit of stellar mass may be passed with the help of positive radial and transverse pressures of the effective $f(T)$ \textit{fluid}.
However, there still exist upper bounds of stellar mass for given model parameters due to the causal condition, i.e.,
the speed of sound of the stellar matter must be less or equal to the speed of light.
For a negative $\alpha$ with sufficient large $|\alpha|$,
this upper bound is apparently greater than the mass limit in GR.
This makes $f(T)$ gravity be able to accommodate some observations of NS that are beyond GR's prediction.
Therefore, we compare the NS models in $f(T)$ gravity with the astrophysical observations of the NS PSR J0030+0451 \cite{Miller:2019cac}, PSR J0740+6620 \cite{NANOGrav:2019jur,Miller:2021qha}, PSR J2215+5135 \cite{Linares:2018ppq},
and the GW event GW170817 \cite{LIGOScientific:2018cki} and GW190814 \cite{Abbott:2020khf}.
For SLy EOS, $\alpha<-0.594r_g^2$ is required to account for all the considered NS observation datasets.
For BSk19 EOS, the models with $\alpha>-6.476r_g^2$ are ruled out by the NS observations.
For BSk20 and BSk21, the observation datasets constrain that $\alpha<-4.4r_g^2$ and $\alpha-2.12r_g^2$, respectively.

These constraints, however, may still be flawed in that an NS model for all of the four EOSs with a negative $\alpha$ can produce large enough stellar mass
only if the central density $\rho_c$ is set to be high enough and gets close to the causal limit.
The EOSs all have their effective ranges and the matter may not be in the same state and phase
when the density and pressure are beyond these ranges.
Moreover, concerning causal limit, a parametrization of EOSs at high density region may be taken into account \cite{Sotani:2017pfj,Astashenok:2021xpm,Astashenok:2021peo}.
Therefore, a joint constraint for both the EOS and the $f(T)$ theory may be more physical.
Another issue may come from the GW.
The constraints of NS parameters from GW events are adapted to the waveform in GR.
A GW theory in $f(T)$ gravity may result in shifts of these constraints.
Finally, the NS models considered in the present work are restricted to static spherically symmetric cases with an isotropic perfect fluid.
It is known that the anisotropy of the fluids and the rotation of a star may raise the stellar mass and reduce the central density \cite{Olmo:2019flu,Astashenok:2020qds}.
These issues are worth further studying in the future.

\section*{Acknowledgement}
\label{ackn}
This work is supported by the National Science Foundation of China under Grant No. 12105179.

\appendix
\section{Parameters $a_i$ in the analytic forms of the EOSs}
\label{appeos}

For SLy EOS, the parameters $a_i$ are listed in Table~\ref{slyparas}.
\begin{table}[h!]
    \centering
    \caption{$a_i$ (SLy) parameters for Eq. \eqref{sly}.}
    \label{slyparas}
    \begin{tabular}{ll|ll}
        \hline
        \hline
        $i$\hspace*{1cm} & $a_i$\hspace*{1.5cm} & $i$\hspace*{1cm} & $a_i$\hspace*{1.5cm} \\
        \hline
        1                & 6.22                 & 10               & 11.4950              \\
        2                & 6.121                & 11               & -22.775              \\
        3                & 0.005925             & 12               & 1.5707               \\
        4                & 0.16326              & 13               & 4.3                  \\
        5                & 6.48                 & 14               & 14.08                \\
        6                & 11.4971              & 15               & 27.80                \\
        7                & 19.105               & 16               & -1.653               \\
        8                & 0.8938               & 17               & 1.50                 \\
        9                & 6.54                 & 18               & 14.67                \\
        \hline
        \hline
    \end{tabular}
\end{table}

For BSk family of EOSs, the parameters $a_i$ in Eq. \eqref{bsk} are listed in Table~\ref{bskparas}.
\begin{table}[h!]
    \centering
    \caption{$a_i$ parameters for BSk family of EOSs \eqref{bsk}.}
    \label{bskparas}
    \begin{tabular}{llll}
        \hline
        \hline
                         & \multicolumn{3}{c}{$a_i$}                                           \\
        \hline
        $i$\hspace*{1cm} & BSk19\hspace*{1cm}        & BSk20\hspace*{1cm} & BSk21\hspace*{1cm} \\
        \hline
        1                & $3.916$                   & 4.078              & 4.857              \\
        2                & 7.701                     & 7.587              & 6.981              \\
        3                & 0.00858                   & 0.00839            & 0.00706            \\
        4                & 0.22114                   & 0.21695            & 0.19351            \\
        5                & 3.269                     & 3.614              & 4.085              \\
        6                & 11.964                    & 11.942             & 12.065             \\
        7                & 13.349                    & 13.751             & 10.521             \\
        8                & 1.3683                    & 1.3373             & 1.5905             \\
        9                & 3.254                     & 3.606              & 4.104              \\
        10               & -12.953                   & - 22.996           & -28.726            \\
        11               & 0.9237                    & 1.6229             & 2.0845             \\
        12               & 6.20                      & 4.88               & 4.89               \\
        13               & 14.383                    & 14.274             & 14.302             \\
        14               & 16.693                    & 23.560             & 22.881             \\
        15               & -1.0514                   & -1.5564            & -1.7690            \\
        16               & 2.486                     & 2.095              & 0.989              \\
        17               & 15.362                    & 15.294             & 15.313             \\
        18               & 0.085                     & 0.084              & 0.091              \\
        19               & 6.23                      & 6.36               & 4.68               \\
        20               & 11.68                     & 11.67              & 11.65              \\
        21               & -0.029                    & -0.042             & -0.086             \\
        22               & 20.1                      & 14.8               & 10.0               \\
        23               & 14.19                     & 14.18              & 14.15              \\
        \hline
        \hline
    \end{tabular}
\end{table}

\section{The critical configurations of NS models}
\label{appcri}
For positive $\alpha$ or for negative $\alpha$ with relatively small $|\alpha|$,
the $M-\mathcal R$ curves of the NS models will have a peak,
corresponding to the critical configurations,
which are listed in Table \ref{tabcri}
and are already marked with circle symbols in Fig. \ref{figMR}.
\begin{table}[h!]
    \centering
    \caption{Some critical configurations of NS models with BSk family and SLy EOSs.}
    \label{tabcri}
    \begin{tabular}{llcll}
        \hline
        \hline
        EOS                      & $\alpha/r_g^2$\hspace*{0.2cm} & $\rho_c/(10^{15}\text{g/cm}^3)$\hspace*{0.2cm} & $\mathcal R/\text{km}$\hspace*{0.2cm} & $M/M_\odot$ \\
        \hline
        \multirow{6}{1cm}{BSk19} & $-0.2$                        & $3.71$                                         & $9.03$                                & $1.89$      \\
                                 & $0$(GR)                       & $3.48$                                         & $9.11$                                & $1.86$      \\
                                 & $0.2$                         & $3.27$                                         & $9.18$                                & $1.84$      \\
                                 & $1$                           & $2.78$                                         & $9.37$                                & $1.75$      \\
                                 & $5$                           & $1.91$                                         & $9.75$                                & $1.48$      \\
                                 & $10$                          & $1.54$                                         & $9.91$                                & $1.29$      \\
        \hline
        \multirow{6}{1cm}{BSk20} & $-0.2$                        & $2.83$                                         & $10.12$                               & $2.19$      \\
                                 & $0$(GR)                       & $2.69$                                         & $10.17$                               & $2.17$      \\
                                 & $0.2$                         & $2.57$                                         & $10.22$                               & $2.14$      \\
                                 & $1$                           & $2.23$                                         & $10.37$                               & $2.05$      \\
                                 & $5$                           & $1.58$                                         & $10.66$                               & $1.76$      \\
                                 & $10$                          & $1.29$                                         & $10.77$                               & $1.53$      \\
        \hline
        \multirow{6}{1cm}{BSk21} & $-0.2$                        & $2.38$                                         & $10.99$                               & $2.30$      \\
                                 & $0$(GR)                       & $2.30$                                         & $11.03$                               & $2.28$      \\
                                 & $0.2$                         & $2.19$                                         & $11.10$                               & $2.26$      \\
                                 & $1$                           & $1.97$                                         & $11.21$                               & $2.19$      \\
                                 & $5$                           & $1.43$                                         & $11.49$                               & $1.93$      \\
                                 & $10$                          & $1.17$                                         & $11.59$                               & $1.71$      \\
        \hline
        \multirow{6}{1cm}{SLy}   & $-0.2$                        & $3.03$                                         & $9.91$                                & $2.07$      \\
                                 & $0$(GR)                       & $2.85$                                         & $9.98$                                & $2.05$      \\
                                 & $0.2$                         & $2.76$                                         & $10.01$                               & $2.03$      \\
                                 & $1$                           & $2.38$                                         & $10.18$                               & $1.95$      \\
                                 & $5$                           & $1.67$                                         & $10.53$                               & $1.68$      \\
                                 & $10$                          & $1.35$                                         & $10.68$                               & $1.47$      \\
        \hline
    \end{tabular}
\end{table}
\bibliographystyle{spphys}
\bibliography{ref}

\end{document}